\documentclass[useAMS,usenatbib]{mn2e}
\usepackage{times}
\usepackage{graphicx}
\usepackage{amssymb}
\usepackage{natbib}
\usepackage{psfig}
\title[Mapping the galaxy NGC 4486]{Mapping the galaxy NGC 4486 (M87) through its Globular Cluster System}
\author[Forte et al.]{Juan C. Forte$^{1,2}$\thanks{E-mail:forte@fcaglp.unlp.edu.ar}, E. Irene Vega$^{1,2}$ and Favio Faifer$^{1,2,3}$ \\
$^1${Facultad de Ciencias Astron\'omicas y Geof\'isicas, Universidad Nacional de La Plata}\\
$^2${Consejo Nacional de Investigaciones Cient\'ificas y T\'ecnicas, Rep.
Argentina}\\
$^3${IALP}\\
}
\begin{document}

\date{Accepted ; Received ; in original form }
\pagerange{\pageref{firstpage}--\pageref{lastpage}} \pubyear{}
\maketitle

\label{firstpage}

\begin{abstract}
As shown in previous works, globular clusters can be used to trace the 
overall structure of the diffuse stellar populations in early type galaxies 
if the number of clusters per unit stellar mass depends on metallicity. 
In this paper we further test this assumption in the galaxy NGC 4486 (M 87), by combining several 
data sources. The results show that globular clusters allow the mapping of the galaxy 
in terms of the surface brightness profile, integrated colour gradient, chemical 
abundance, and  mass to luminosity ratios up to 1000 arcsec (or 80.4 kpc) from its centre (i.e. some 10 
effective radii). The analysis indicates the presence of a dominant high metallicity bulge 
associated with the red globulars, whose ellipticity increases outwards, and of a more flattened 
low metallicity halo connected with the blue globulars. The chemical abundance gradient of 
the composite stellar population is remarkably similar to that inferred from X ray observations 
of hot gas. The mass-metallicity spectrum of the stellar population can, in principle, be understood in 
terms of inhomogeneous enrichment models. In turn, the distribution of the bluest GCs, and lowest 
metallicity halo stars, has an intriguing similarity with that of dark matter, a feature  shared 
with NGC 1399. Also, in these two galaxies, the number of blue GCs per dark mass unit is 
identical within the errors, $\approx $ $1.0(\pm 0.3) \times 10^{-9 }$. The total stellar 
mass derived for NGC 4486 is $6.8(\pm 1.1)\times10^{11} M_\odot$ with a baryonic mass 
fraction $f_{b}=$ 0.08($\pm$ 0.01). 

\end{abstract}

\section{Introduction}
\label{INTRO}

 The potential of globular clusters (GCs) as tracers of early events
 in the  life of galaxies has been emphasized along the years. The idea
 has roots in \citet{EGG62} and \citet{SEA78} and a landscape of the different
 issues within that context is given, for example, in \citet{BRO06}. 
 The theoretical aspects of the problem have been addressed in an effort to define its
 cosmological background (e.g. \citealt{BEA02}; \citealt{BEK08}; \citealt{MURG10}). 

 However, as noted by \citet{KPA09}, and also see \citet{SPIT10}, some important
 aspects remain as open questions. Among them, the striking differences observed between
 the metallicity distribution function (MDF) of the CGs, usually bimodal, and that of field
 stars as observed in resolved galaxies. In this last case the MDFs are predominantly broad,
 asymmetric, and  exhibit a low  metallicity tail (e.g. \citealt{DUR01}; \citealt{HAR02}; 
  \citealt{REJ05}; \citealt{REJ11}).

  Differences and similarities between GCs and the underlying stellar populations have been
  pointed out in the past. For example, \citet{FOR81} and \citet{STR81} found that GCs
  appear significantly bluer than the stellar haloes at all galactocentric distances in four
  Virgo ellipticals.

  These comparisons should take into account, however, that the CGs statistics are
  ``number weighted'' while halo colours are, naturally, ``luminosity weighted''. Thus, a
  coincidence between GC and stellar halo colours would be expected only if each GC formed along
  a given diffuse stellar mass on a constant luminosity basis. This fact is frequently ignored
  in the literature.

  A possible quantitative connection between the features of  GCs and field stars, assumes
  that clusters are born during major star forming episodes and that, the number of GCs
  per diffuse stellar mass unit, increases exponentially when the logarithmic chemical
  abundance $[Z/H]$ decreases.  This connection is consistent with the variation of the GCs 
  specific frequency $S_{n}$ with chemical abundance (\citealt{HAR02}), and allows the recovery
  of both the luminosity profile and colour gradient in a galaxy as the result of a multi-metallicity
  stellar population.

   A first approach along these lines was presented in  \citet{FOR07}
  (hereafter FFG07). More consequences of that idea were discussed in \citet{FOR09}
  (hereafter FVF09) who performed an analysis of 69 galaxies within the Virgo ACS (see
  \citealt{FER04}; \citealt{PEN08}).

  FVF09 also find that globular cluster systems (GCS) in Virgo galaxies with stellar masses
  below $M_*\approx 10^{11} M_\odot$, lie on plane defined (in a 3-D logarithmic space) by GC formation
  efficiency (number of GCs per  unit stellar mass), total stellar mass and projected stellar mass
  density. NGC 4486, and other bright Virgo ellipticals, lie well outside this plane, probably reflecting
  the effects of past merger events. 

  In this paper we revise the case of NGC 4486 (M 87), taking advantage of a number
  of photometric works (\citealt{TAM06}; \citealt{HAR09}; \citealt{JOR09}) that, combined, allow an 
  analysis on an angular scale of $\approx$ 1000 arcsec.
  
  An important assumption of the analysis is that GCs and field stars are  
  mostly coeval or formed  within a narrow period of time. Arguments in 
  support of this hypothesis can be found, for example, in 
  \citet{SPIT10} and references therein. In particular, a direct estimate
  of the age of the halo of the galaxy NGC 3923, as well as of its associated GCs,
   was presented by \citet{NOR08} using Gemini MOS spectra and Lick
  indices. These results also indicate and old age (about 12 Gyr) for both stars and GCs
  (see \citealt{BEA08} for the case of NGC 5128).
 
  More arguments in favour of old ages and multi-metallicity spectra for the dominant
  stellar populations in elliptical galaxies, stand on the analysis of their integrated colours
  as discussed in \citet{SCH09}.

  Several papers, on the other side, deal with the surface brightness profile of NGC 4486.
  In this work we adopt the blue band (Johnson system)
  surface brightness profile presented  by \citet{CAO90}. All these works denote the existence of
  a flattened cD halo (already detected by \citealt{ARP71}), traced to very low
  luminosity levels in \citet{MIH05}. More  recently, \citet{RUD10} find that the $(B-V)$
  colour gradient of the galaxy has a logarithmic dependence along the semi major axis over an
  angular range of (at least) $1000$ arcsec (i.e., approximately, ten effective radii). This
  corresponds to 80.1 kpc for a distance modulus $(Vo-M_v)=31.1$, or 16.5 Mpc, (\citealt{TON01}),
  adopted in this work .
 
  The paper is organized as follows: Section 2 presents an improved version of the
  GC integrated colours vs. metallicity relation given in FFG07 and the
  connection between chemical abundance Z (in solar units hereafter) and the cluster integrated
  colours, including an attempt to model the so called ``blue tilt''. Section 3 explains the
  assumed link  between GCs and field stars. Section 4 describes the fit of the GC colour histograms
  within the inner $300$ arcsec  along the galaxy semi major axis, aiming at determining
  the number of clusters belonging to the blue or red sub-population as well as their 
  respective chemical scale-lengths. A discussion of the large angular scale features of 
  the galaxy, including the shape of the surface brightness profile, colour gradient, stellar
  metallicity function and stellar mass to luminosity ratios is given in Section 5. The 
  inferred halo, bulge and total stellar masses, and a comparison with the dark matter content 
  is presented in Section 6. Finally, the GCs formation efficiencies for both the blue and 
  red GCs and the conclusions of this paper, are included in Section 7 and 8, respectively.

\section{The globular clusters colour distribution.}
\label{HISTOS}

 In their analysis of the GCS systems associated with Virgo ellipticals, \citet{PEN06} present
 non parametric fits of the CG colour histograms. In this paper we adopt  a different approach, that
 seeks the determination of the chemical abundance scales as well as the fraction of clusters in each 
 population, on the basis of a Monte Carlo simulation. 

 As discussed in FFG07, the simplest function that links the chemical abundance Z  with the integrated
 $(C-T_1)_o$ colours through the empirical calibration, and within the inherent statistical noise of
 the GC counts, is

\begin{equation}
dN(Z)/dZ  {\approx} \exp({-{(Z-Zi)/Z_{s})}}\
\end{equation}

\noindent  where $Zi$ is a pedestal abundance  and $Z_{s}$ is the characteristic abundance
 scale length of each GC sub-population. 

 The modelling of the GCs colour bimodality, is then similar to that given in FFG07. First, a ``seed''
 globular is randomly generated with a given Z governed by the $ Z_{s}$parameter. The
 synthetic integrated colours were in turn obtained from the empirical colour  metallicity
 calibration and adding Gaussian observational errors. A colour excess $E(B-V) = 0.02$ from
 \citet{SCH99}, that transforms to $E(C-T_1) = 0.04$, was also added to the intrinsic $(C-T_1)_o$
 colours in order to match the FFG07 photometry. $T_1$ magnitudes were also generated for each
 cluster adopting a GC Gaussian luminosity function with a turn over at $T_1=23.2$ and characterized
 by a dispersion $\sigma (T_{1})=1.60$ mag. A detailed discussion of the behaviour of this parameter
 for the Virgo ACS galaxies is given in \citet{VIL10}.
 
 The shape of the empirical colour-metallicity relation requires a GC bimodal metallicity
 distribution in order to match the observed colour bimodality defined in terms of the
 so called blue and red GCs (for which we adopt upper chemical abundance values $Z=Z_\odot$ and
 $Z=3.5 Z_\odot$, respectively). Previous, and more recent arguments, in favour of  bimodal GC metallicity
 distributions can be found, for example, in \citet{BRI11} (however, see \citealt{YOO11} for a different
 interpretation of the colour bimodality).

 FFG07 presented an empirical colour-metallicity relation based on 100 MW GCs with $(C-T_1)$ Washington sytem
 colours and 98 extragalactic GCs with $(g'-i')$ colours in the (modified) Gunn system. In turn, \citet{MOY11}
 derived an improved version of that calibration:
\begin{eqnarray}
(C - T_1)_{o} = 0.91(\pm 0.15)+0.06(\pm0.02)([Fe/H]_{zw} + \nonumber\\
 3.70(\pm0.87))^2
\end{eqnarray}

\noindent where $[Fe/H]_{zw}$ are metallicities on the \citet{ZIN84} scale. The improvement comes from
 a better determination of the relation used to transform $(g'-i')_o$ to $(C-T_1)_o$:

\begin{equation}
(C - T_1)_{o} = 1.72(\pm 0.04)~(g'-i')_{o} - 0.07 (\pm 0.06)
\end{equation}

\noindent  obtained by observing a reference field in NGC 4486 with Gemini/MOS (Forte et al. in preparation)
 with GC candidates observed in both photometric systems. In turn, from FFG07 and \citet{JOR09}:

\begin{equation}
(C - T_1)_{o}= 1.26(\pm 0.04)~(g-z)_{o, ACS} + 0.01(\pm 0.05)
\end{equation}

Other colour-colour relations used in this work are:

\begin{equation}
(B-V)_{o}= 0.51(\pm 0.03)~(C - T_1)_{o} + 0.06 (\pm 0.04)
\end{equation}

\begin{equation}
(B-R)_{o} = 0.70(\pm 0.06)~(C - T_1)_{o} + 0.27(\pm 0.07)
\end{equation}

\begin{equation}
(V-I)_{o}= 0.50(\pm 0.06)~(C - T_1)_{o} + 0.32(\pm 0.07)
\end{equation}

\noindent derived from observations of galactic globulars (see, also \citealt{FORB01}) and can be
used to transform the relation given in eq. 2 to other colours.

 A similar calibration based on 40 MW GCs, and presented in \citet{SIN11}, shows very good agreement
 with ours in the range defined between $[Fe/H]$ $-2.5$ and $-0.5$. At higher metallicities, their calibration
 (that includes 3 MW GCs, compared with 26 GCs that define eq. 2) becomes steeper and leads to $(C - T_1)$ colours
 redder by 0.15 mag at $[Fe/H]$ = 0.0.
  
 An analysis of the colour residuals from our adopted calibration as a function  
 of the GC ages, determined from colour magnitude diagrams for the MW GCs, or from Lick  indices for
 the extragalactic globulars, reveals no systematic trends. The empirical calibration, then, corresponds
 to the mean age of the MW GCs, that we assume as $12$ $Gyr$.

 For any other age $\tau$ (in $Gyr$) the colour transforms to:

\begin{equation}
(C - T_1)_{\tau} = (C - T_1)_{o} + \Delta(C-T_1)
\end{equation}

\noindent with
  
\begin{equation}
\Delta(C-T_1)=(\tau-12.0) \cdot d(C-T_1)/d\tau
\end{equation}

\noindent where the term $d(C-T_1)$/d$\tau$ can be approximated, for ages between $8$ and $15$ $Gyr$,
 as:

\begin{equation}
d(C-T_1)/d\tau=0.02 + 5.3\times10^{-3} \cdot [Fe/H]_{zw}
\end{equation}
 
\noindent derived from  the single stellar population models by \citet{BRU03}.

On the other side, \citet{MEN07} connect the $[Fe/H]_{zw}$ metallicity scale with
the total chemical abundance of the \citet{THO04} models through:

\begin{equation}
[Z/H] = [Fe/H]_{zw} + 0.131       
\end{equation}

 The GCs colour magnitude diagram  for NGC 4486 shows the so 
 called ``blue tilt'' (\citealt{PE09}, and references therein), a feature 
 explained in FFG07 as an increase of the $Zi$ parameter with the GC brightness 
 and mass. Alternative views about the interpretation of the blue tilt, and the role of 
 chemical pre and/or self enrichment, are given in \citet{STR08} and \citet{BAI09}.

 A linear relation between colours and GC magnitudes was adopted in FFG07 to describe the
 tilt. However, a new analysis rather suggests a non linear trend (see, also, the colour
 magnitude diagrams presented by \citealt{HAR09}):

\begin{equation}
(C-T_1)_{t}=21.67\cdot T_{1}^{-0.95}
\end{equation}

\noindent  where $(C-T_1)_{t}$ and $ T_{1}$ are (reddening corrected) colours and magnitudes
 that define the tilt shape. At a given  $T_{1}$, the $(C-T_1)_{t}$ colour  leads to
 a metallicity defined through the empirical colour-metallicity relation,
$[Fe/H]_{t}$, and then to a $[Z/H]_{t}$ from the Mendel et al. relation:

\begin{equation}
[Z/H]_{t}= [Fe/H]_{t} + 0.131
\end{equation}

 In this paper, we attempt modelling the chemical variation associated with the
 tilt by defining:

\begin{equation}
Zi = Zoi + \Delta Z
\end{equation}

\noindent where $Zoi$ is a fit parameter (the lowest chemical abundance for a given
GC sample), and $\Delta Z$ is given by: 

\begin{equation}
\Delta Z = 10^{[Z/H]_{t} } - 0.092
\end{equation}

 This procedure leads to values that range from  $\Delta Z$ = 0.06 at $T_{1}=19.0$ mag, 
 corresponding to the brightest objects, to $\Delta Z$ = 0.0 at  $T_{1} = 23.2$ mag, the
 turn-over of the GCs integrated luminosity function, where the tilt seems to disappear.

\begin{table*}

\centering
 \begin{minipage}{150mm}
  \caption{Fit parameters for the GC Colour Histograms, Colours and Chemical Abundances for GCs and NGC 4486}

\begin{tabular}{@{}cccccccccccccc@{}}
\hline
  $\bar{a}$ & $\epsilon$ & $N_{blue}$ & $Z_{SB}$ & $Z_{oiB}$
     & $[Z/H]$ & $N_{R}$ & $Z_{SR}$ & $Z_{oiR}$ & $[Z/H]$ 
     & $(B-R)$ & $(B-R)$ & $[Z/H]_M$  & $[Z/H]_L$ \\
      ($\arcsec)$ & & & & &BGC& & & &RGC &GCs & Galaxy & Galaxy & Galaxy \\

 \hline
 18.7 & 0.03 &  55 & 0.085 & $10^{-3}$ & -1.09 &   81  &  1.45  &  0.10  &  -0.02  &  1.40  &  1.56  &  0.21 &  0.12\\
 45.5 & 0.07 & 105 & 0.030 & $10^{-3}$ & -1.41 &  185  &  1.35  &  0.05  &  -0.07  &  1.37  &  1.52  &  0.19 &  0.11\\
 74.7 & 0.10 & 140 & 0.030 & $10^{-3}$ & -1.41 &  140  &  1.35  &  0.05  &  -0.07  &  1.33  &  1.50  &  0.18 &  0.09\\
105.3 & 0.13 & 150 & 0.035 & $10^{-3}$ & -1.37 &  159  &  1.35  &  0.10  &  -0.04  &  1.33  &  1.48  &  0.16 &  0.06\\
135.2 & 0.16 & 119 & 0.035 & $10^{-3}$ & -1.37 &  119  &  1.20  &  0.10  &  -0.06  &  1.32  &  1.47  &  0.14 &  0.04\\
165.5 & 0.18 & 120 & 0.020 & $10^{-3}$ & -1.52 &  130  &  0.80  &  0.05  &  -0.21  &  1.27  &  1.46  &  0.07 & -0.03\\
194.4 & 0.21 & 139 & 0.020 & $10^{-3}$ & -1.52 &  118  &  0.85  &  0.05  &  -0.19  &  1.28  &  1.46  &  0.09 & -0.02\\
224.5 & 0.23 & 130 & 0.025 & $10^{-3}$ & -1.46 &  103  &  0.95  &  0.05  &  -0.16  &  1.26  &  1.45  &  0.08 &  0.00\\
255.4 & 0.25 & 140 & 0.020 & $10^{-3}$ & -1.52 &   95  &  0.85  &  0.05  &  -0.20  &  1.24  &  1.45  &  0.08 & -0.04\\
284.7 & 0.27 & 120 & 0.020 & $10^{-3}$ & -1.52 &   85  &  0.80  &  0.05  &  -0.21  &  1.25  &  1.44  &  0.06 & -0.06\\
\hline
\end{tabular}
{col. 1: Mean {\bf a} semi major axis of the annulus; col.2: Mean ellipticity; col.3: Number of blue GCs; col.4: Chemical
scale length for the blue GCs; col.5: Initial chemical abundance for blue GCs; col.6: Mean chemical
abundance for the blue GCs; col.7: Number of red GCs; col.8: Chemical scale length for red GCs; col.9:
initial chemical abundance for red GCs; col.10: Mean chemical abundance for red GCs; col. 11: Mean (B-R)
colours for the whole GC population; col.12: Integrated (B-R) colour of the galaxy; col. 13: Mean chemical
abundance for the galaxy (mass weighted); col. 14: Mean chemical abundance for the galaxy (B luminosity weighted)}
\end{minipage}
\label{Table1}
\end{table*}
\section{The globular clusters-field stars connection.}
\label{GLOBSTARS}
 As shown in several papers,  the metallicity distributions
 of GCs and resolved field stars are noticeably different, e.g., \citet{DUR01} in
 M 31 or \citet{REJ11} in NGC 5128. The tentative link between these
 systems, explored in FFG07 and FFG09, comes from the fact that both the number $N$ of
 GCs per unit stellar  mass (e.g., \citealt{MIL07}) and the metallicity of a galaxy (\citealt{TRE04})
 can be approximated as power laws of the stellar galaxy mass. Combining these relations, and
 adding an explicit dependence with chemical abundance, leads to: 

\begin{equation}
dN(Z)/dM_*(Z) = {\gamma} \exp({-{\delta [Z/H])}}\
\end{equation}

 This equation connects  $dN(Z)$, derived from the GC colour histograms, to an (assumed coeval) diffuse
 stellar mass element $dM_*(Z)$, whose mass to luminosity ratio and colour solely depend on 
 chemical abundance.

 Then, the knowledge of the projected GC areal density at a given galactocentric radius, allows the fit
 of the colour and surface brightness of the galaxy by iterating the $\delta$ and $\gamma$
 parameters, and integrating in the chemical abundance domain covered by the GCs. Eq. 16 imply that 
 the integrated colours, composite $(M/L)$ ratios and mean chemical abundances (mass or luminosity
 weighted) are governed by $\delta$ since this parameter regulates the relative masses of stars
 associated to GC with different chemical abundances.  
  
 As in FF07 and FVF09, the {\it B} luminosity associated with $M_*(Z)$,
 was obtained from:

\begin{equation}
(M/L)_{B} = 3.71 + ([Z/H] + 2.0)^{2.5} 
\end{equation}

 \noindent which gives a good approximation to the mass to {\it B} luminosity ratios, 
 given by \citet{WOR94} for a Salpeter stellar mass function. An idea about the uncertainty of
 this relation comes from a comparison with other models. For example,  the
 \citet{MAR04} models show an agreement within 10 per cent with Worthey's although, at the lowest
 chemical abundance, this last author gives ratios $\approx$ 24 per cent  larger. Other source
 of uncertainty is connected with the assumed stellar mass function. The adoption
 of a \citet{CHAB03} function leads to stellar masses systematically smaller by a factor of 0.55.

\section{NGC 4486: The inner 300 arcsec (24.1 kpc).}
\label{inner}

 In this section we discuss the inner 300 arcsec along the semi major
 axis of the galaxy, where the photometric works by \citet{JOR09} 
 and FFG07 give a complete areal coverage for objects brighter than the GCs turn-over.
 At this magnitude level  we estimate a background
 contamination of about one object per sq. arcmin, which agrees with the
 value given by \citet{HAR09}, after taking into account the slightly
 different limiting magnitude in both works.

 For this region in particular, the total expected background
 contamination  should be below $5$ per cent, then minimizing the effects of these
 objects on the colour histograms of the GC candidates.

 In contrast with FFG07, where single scale parameters $Z_{SB}$ and $Z_{SR}$ were adopted
 to characterize either the blue or the red GC population, we performed distinct fits within
 elliptical annuli ($30$ arcsec  wide) following the ellipticity variation of the galaxy
 halo in an attempt to detect  chemical abundance gradients. 
 
 An analysis of all the colour histograms shows a small asymmetry in the sense 
 that GCs to the NE of the semi major axis of the galaxy seem slightly redder (by $\approx$ 0.015 in
 $(C-T_{1})$) than those to the SW of that axis, and may indicate differential reddening. Besides,
 the three outermost annuli show a bump at $(C-T_{1}) \approx 0.95$ (see the Appendix).

 The nature of these objects (around fifty in total) remains unclear although we
 noticed that they appear mostly to the NE of the galaxy major semiaxis and
 without any obvious concentration towards the centre of the galaxy. As a tentative
 explanation, they might be foreground MW stars connected with the so called
 ``Virgo over-density'' (see, for example, \citealt{LEE10}).

 The GC colour histograms were fit by iterating the $Z_{oi}$ and $Z_{s}$ parameters as
 well as the number of clusters of each sub-population and minimizing the ``quality fit''
 parameter defined by \citet{COT98}. These values  are listed in Table 1. Typical errors
 of the $Z_{SB}$ and $Z_{SR}$ parameters are $\pm$ 0.005 and $\pm$ 0.05 respectively.

 The observed and model GC colour distributions are displayed in the Appendix, both in discrete
 bins and smoothed histograms.
 
\begin{figure}
\resizebox{1\hsize}{!}{\includegraphics{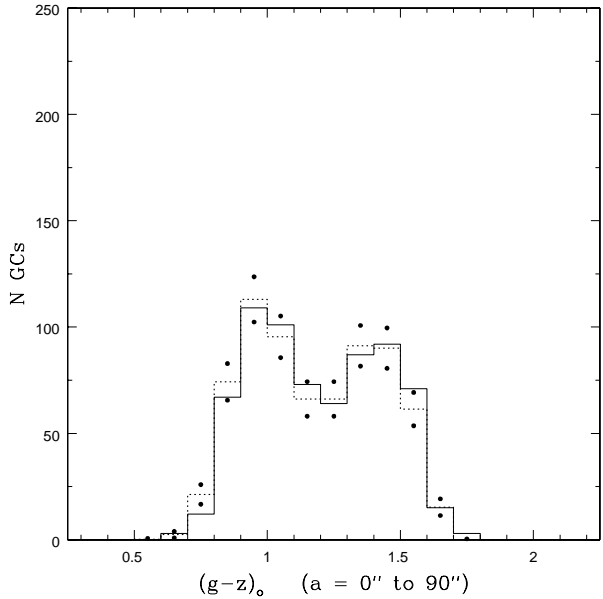}}
\caption{$(g-z)_o$ colour histogram for 706 GC candidates within a
  semi major axis $a=90$ arcsec (7.2 kpc), where the areal coverage given in Jord\'an
  et al. (2009) is complete. The dotted line shows a composite model based on the
  parameters listed in Table 1. Dots indicate the statistical uncertainty of the
  model within each 0.1 mag. bin. 
}
\label{figure1}
\end{figure}

\begin{figure}
\resizebox{1\hsize}{!}{\includegraphics{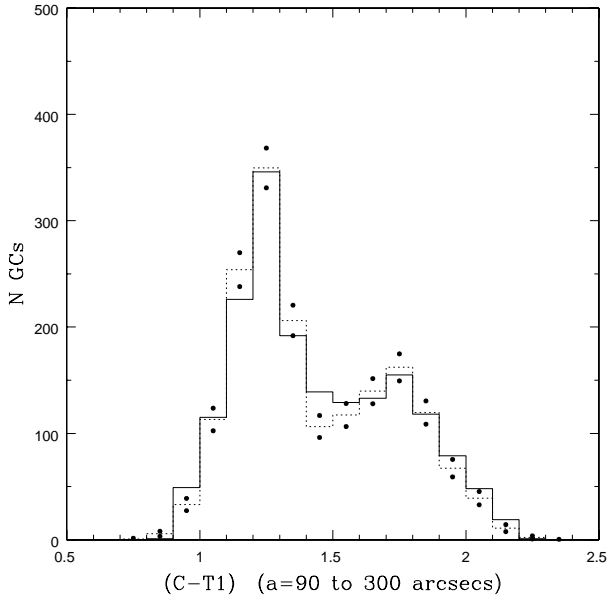}}
\caption{$(C-T_1)$ colour histogram for 1727 GC candidates between $a=90$ and 
 $300$ arcsec (7.2 to 24.1 kpc) where the areal coverage of the photometry by Forte, 
 Faifer and Geisler (2007) is complete. Dotted lines and dots have the same meaning
 as in the previous figure.
}
\label{figure2}
\end{figure}

 The GC colour histograms corresponding to regions defined between
 $a=0$ to $90$ arcsec and $a=90$ to $300$ arcsec are depicted in Figures \ref{figure1} and
 \ref{figure2} . These diagrams also display the {\bf{composite}} model colour histograms obtained
 by combining the individual fits listed in Table 1. This table shows a decrease of the
 GCs chemical scales $Z_{SB}$ and $Z_{SR}$ along the semi major axis of the galaxy, and also a
 difference in the sense that red GCs require higher values of the initial abundances $Z_{oi}$, 
 compared to those of the blue GCs. 

 The $\Delta Z$ component of the $Zi$ parameter (Section 2) was also included
 in the colour modelling of the red GCs. However, their large chemical abundance scale length, 
 compared with the $\Delta Z$ parameter, makes the tilt effect practically undetectable
 for these clusters.

\begin{figure}
\resizebox{1\hsize}{!}{\includegraphics{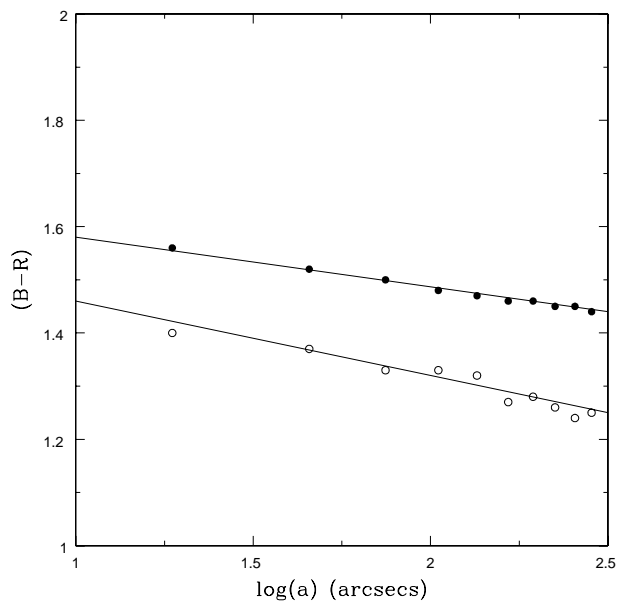}}
\caption{Logarithmic (B-R) colour gradients along the NGC 4486 semi major axis (within $30$ arcsec wide
 elliptical annuli). Galaxy: filled dots. Globular clusters (mean values): open dots.
}
\label{figure3}
\end{figure}

 The $(B-R)$ colour gradient of the galaxy and the mean integrated colours of the {\bf{whole}} cluster population
 are shown in Fig. \ref{figure3}. This last gradient is steeper than that of the galaxy, an
 effect noted for example, by \citet{LIU11}. We stress that both gradients are sensitive to 
 eventual chemical gradients but, while the first one reflects the luminosity weighted contribution 
 of the different diffuse stellar populations, the GC gradient is  sensitive to the changing 
 ratio in number of blue to red clusters along the galaxy major axis.

 The galaxy  gradient in the last figure is based on the {\it BVRI} data discussed by
 \citet{MIC00}. This author determined the sky level in each photometric band through
 an iterative procedure  that led to consistent {\it ``nice logarithmic gradients''} .
 
 Transforming the Michard's colour gradients to $(B-R)$, through the colour-colour relations given in 
 section 2, leads to a mean gradient:\\

 $\Delta(B-R)$/$\Delta$$log(a)=-0.11 \pm 0.02 $.\\

 The $\delta$ parameter that fits the $(B-R)$ colour of the galaxy within each annular region
 was determined using the GCs parameters listed in Table 1. In turn, the $\gamma$ parameter
 was obtained in such a way that the model $B$ magnitude agreed with that obtained by integrating
 the \citet{CAO90} profile within the inner and outer boundaries of each annulus.

\begin{figure}
\resizebox{1\hsize}{!}{\includegraphics{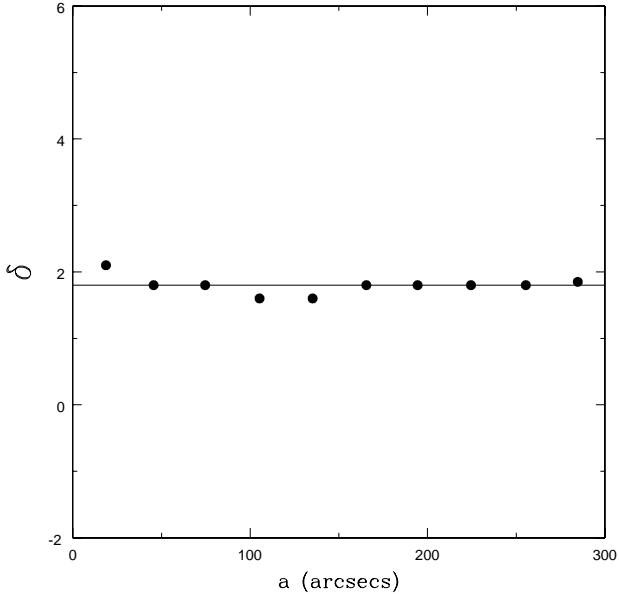}}
\caption{Behaviour of the $\delta$ parameter along the NGC 4486 semi major axis (within $30$ arcsec
 wide elliptical annuli). The horizontal line corresponds to $\delta=1.80$.
}
\label{figure4}
\end{figure}

\begin{figure}
\resizebox{1\hsize}{!}{\includegraphics{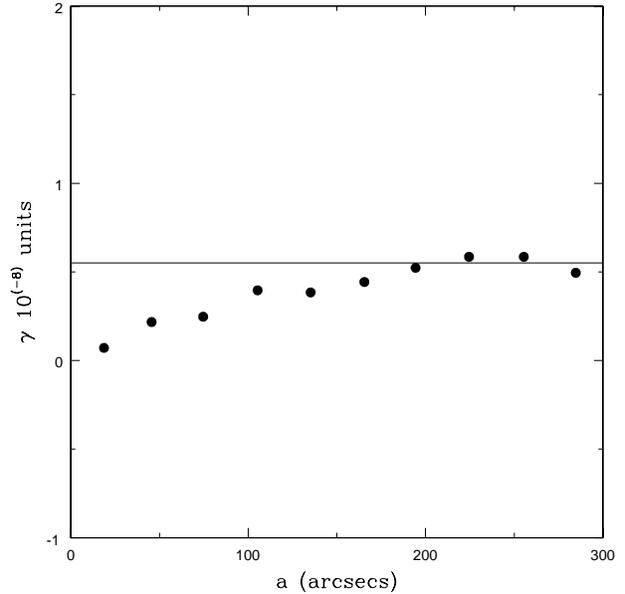}}
\caption{Behaviour of the $\gamma$ parameter along the NGC 4486 semi major axis (within $30$ arcsec wide
 annuli). The sample includes only globulars brighter than the GCs turn-over magnitude, i.e,
 about 50 per cent of the total GC population. The decrease of this parameter towards the galaxy
 centre indicates a smaller number of GCs per unit  of diffuse stellar mass. The horizontal
 line corresponds to $\gamma=0.55\times10^{-8}$.
}
\label{figure5}
\end{figure}

 The run of the $\delta$ and $\gamma$ parameters along semi major axis of the galaxy are displayed in
 Fig. \ref{figure4} and Fig. \ref{figure5} respectively. The first one shows a mean  value of $ 1.80 $ 
 with an rms of $\pm 0.05$. In contrast, $\gamma$ rises in a monotonic way until reaching a value of
 about $0.55 \times 10^{-8}$ at $a\approx$ 250 arcsec . This trend indicates an increasing number of
 GCs per diffuse stellar mass, and may be explained as the result of dynamical
 destruction/erosion effects, which  become less important with increasing galactocentric radius
 (see, for example, \citealt{CAP09} and references therein).



\begin{figure}
\resizebox{1\hsize}{!}{\includegraphics{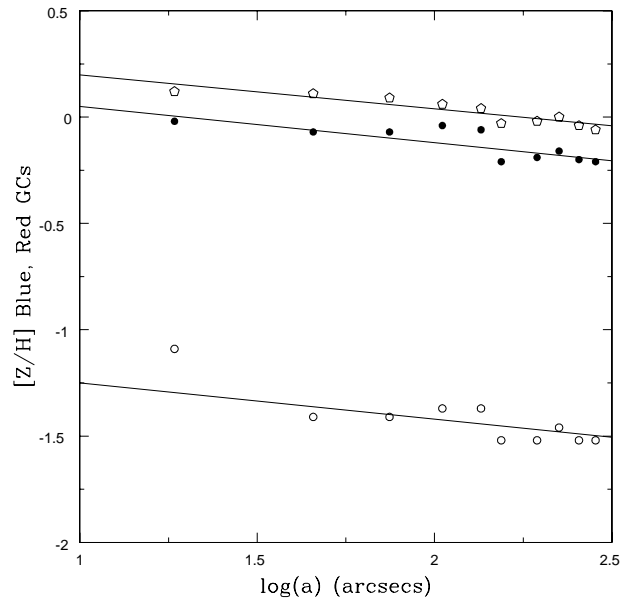}}
\caption{Chemical abundance $[Z/H]$  gradient for the red (filled circles) and blue GCs 
 (open circles), respectively (number weighted). Open pentagons represent the blue
 luminosity weighted abundance of the galaxy. The straight lines have a slope
 $\Delta[Z/H]$/$\Delta$$log(a)=-0.17$. 
}
\label{figure6}
\end{figure}


\begin{figure}
\resizebox{1\hsize}{!}{\includegraphics{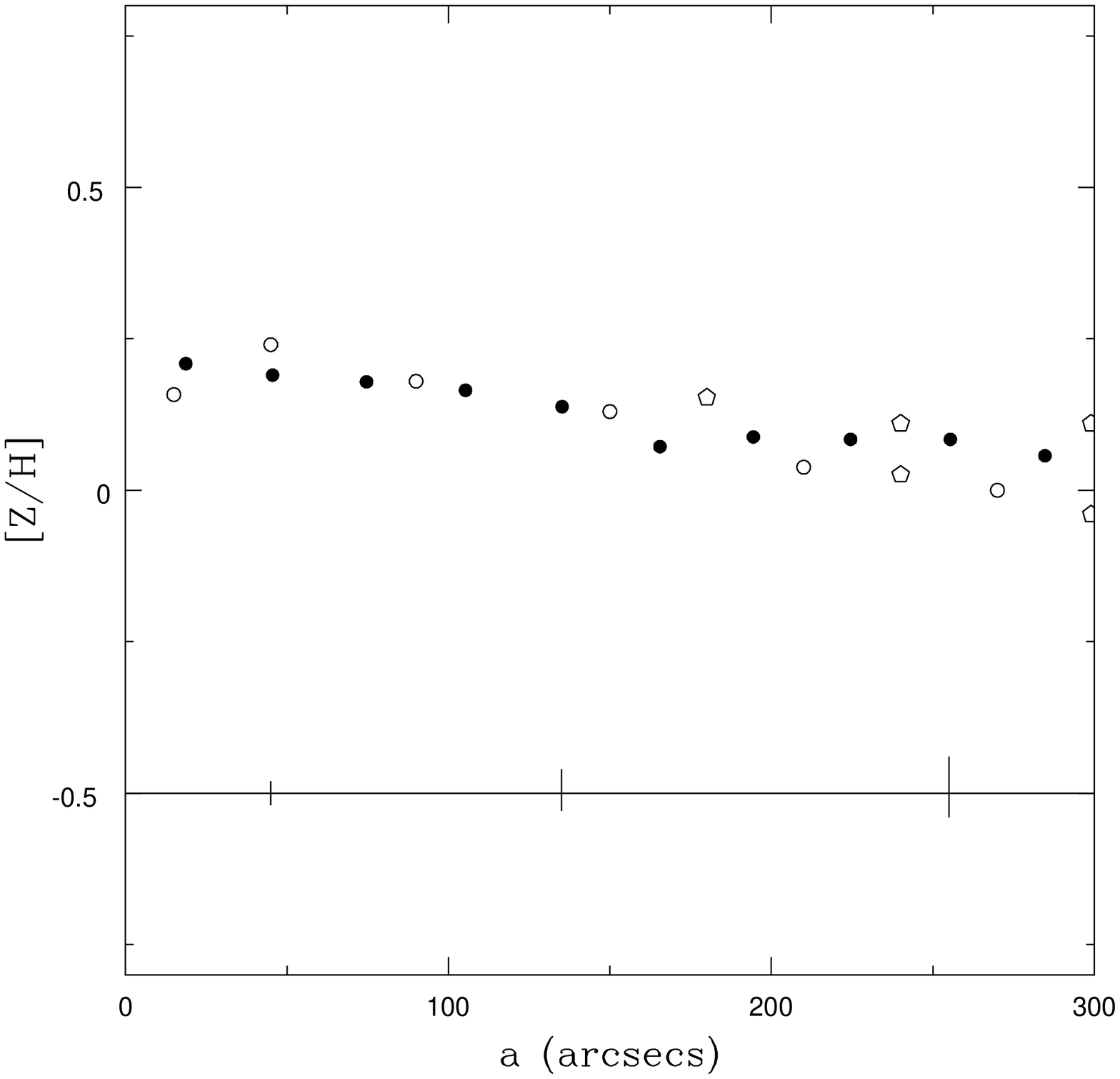}}
\caption{Chemical abundance $[Z/H]$  inferred for the NGC 4486 stellar halo
 along its semi major axis (solid dots). Open dots and pentagons are Fe are abundances
 from X ray  observations  given in \citet{GAS02} and \citet{SIM10} respectively.
 The Fe abundances from these two works have been  shifted upwards by 0.18.
 The vertical lines at the bottom show the effect of increasing/decreasing the
 adopted age by $\pm$ 1.5 Gyr, then decreasing/increasing the inferred abundances
 (see text).
}
\label{figure7}
\end{figure}

 Table 1 also includes the  mean chemical abundance (number weighted) for the blue and red GCs ,
 and the mean chemical abundance  (mass and luminosity weighted)  of the inferred composite stellar
 population. Linear least square fits to the blue GC abundances (with $a$ in arcsec) yield:

\begin{equation}
[Z/H]  = -0.17(\pm 0.07)~log(a)-1.08(\pm 0.15)      
\end{equation}
\noindent and for the red GCs:
\begin{equation}
[Z/H]  = -0.17(\pm 0.05)~log(a)+0.22(\pm 0.09)      
\end{equation}
\noindent in turn, for the galaxy (mass weighted): 
\begin{equation}
[Z/H]  = -0.14(\pm 0.02)~log(a)+0.41(\pm 0.04)      
\end{equation}
\noindent and (blue luminosy weighted): 
\begin{equation}
[Z/H]  = -0.16(\pm 0.03)~log(a) +0.36(\pm 0.05)      
\end{equation}

 The  gradient fits corresponding to the GCs and to the galaxy (B luminosity
 weighted), are depicted in Figure \ref{figure6}.

 Excluding the [Z/H] value of the innermost blue clusters, the gradients of both GC
 sub-populations seem rather similar. However, taking into account the uncertainties
 of the derived coefficients, we cannot dismiss a lower or even null colour 
 gradient for the blue GCs. 

 To within the uncertainties, our result is in very good agreement with \citet{HAR09}
 who finds  {\bf{galactocentric}} slopes of $-0.12(\pm 0.02)$ and $-0.17(\pm 0.03)$ for
 the blue and red GCs respectively.

 The significance of the $[Z/H]$ value corresponding to the blue GCs in the innermost annulus  
 would deserve further analysis since, if confirmed by other means (e.g. spectroscopy), would imply
 an enhanced chemical enrichment process, probably at the very early stages of the formation of the galaxy.    

 Figure \ref{figure6} also shows another particular feature: the abundance residuals from the linear
 fits seem to occur in phase for both the blue and red GCs along the semi major axis of the galaxy. A possible
 explanation to this effect would be differential interstellar reddening affecting the colours from
 which metallicities are derived. Such a kind of GC colour fluctuations are in fact seen in Fig. \ref{figure3}
 (but not on the galaxy colours, as they come from a linear fit to the data in Michard's work).  

 In turn, Fig. \ref{figure7} shows the (mass weighted) mean chemical abundance of the stellar population
 along the semi major axis of the galaxy, as well as the Fe abundances determined by \citet{GAS02} and
 \citet{SIM10} on the basis of hot gas X ray observations (both set of observations were shifted upwards
 by 0.18 in ordinates). Stars and hot gas  behave in a remarkably similar fashion suggesting a strong coupling.

  The effects of changing the adopted age ($12$ Gyr) of the system on the chemical abundances, and following the
  prescription described in Section 2, are also shown  in  Fig. \ref{figure7}. As an example, the vertical bars correspond 
  to an increase/decrease of  1.5 Gyr, leading to abundances that decrease/increase within a range of $-0.05$ to $+0.10$
  in $[Z/H]$. The assumption of an older age would not have an impact on the overall gradient while a younger
  age would lead to an even shallower gradient. 

\begin{figure}
\resizebox{1\hsize}{!}{\includegraphics{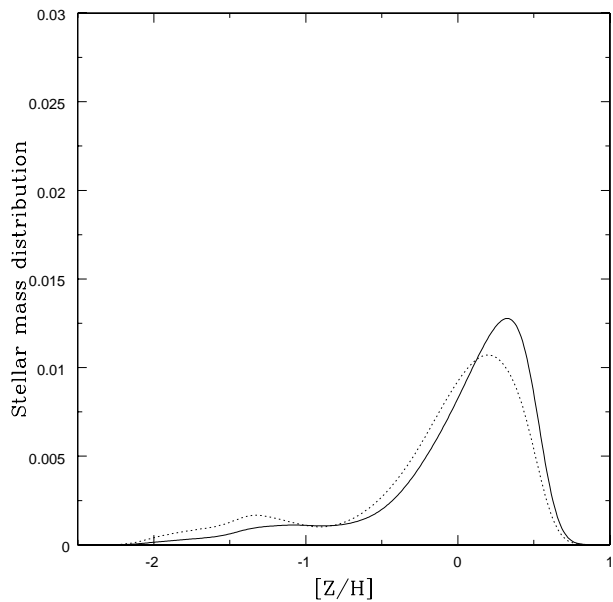}}
\caption{Inferred stellar metallicity distribution function  (normalized by total mass in units of relative mass per
 0.01 dex in metallicity) for stars within two different ranges along the NGC 4486 semi major axis: 100 arcsec ($\approx$
 one effective radius), solid line; and $100$ to $300$ arcsec.  These distributions show the effect of the chemical
 gradient and the presence of a low metallicity tail that becomes more evident in the outer region.  
}
\label{figure8}
\end{figure}

 The stellar mass statistics inferred from eq. 16, corresponding to $\delta=1.80$, and convolved with a Gaussian kernel
 ($\sigma=0.1$), is shown in Fig. \ref{figure8} for two different ranges along the semi major axis: within one
 effective radius ($\approx$ 100 arcsec or 8 kpc); \citealt{CAO90} and within one to three times that value.

 These distributions show a broad high abundance component with a peak that shifts towards smaller values with increasing
 galactocentric radius while the presence of a low metallicity tail, reaching $[Z/H]\approx$ $-2.0$, becomes more prominent.

 Our inferred stellar mass-chemical abundance distribution has a strong similarity with the MDFs of resolved stars in NGC 5128 
 (\citealt{REJ05}; \citealt{REJ11}).  From their analysis, these authors conclude that the bulk of the stars in NGC 5128
 are 11 to 12 $Gyr$ old and exhibit a range in chemical abundance that spans from $(Z/Z_\odot)$$\approx$ 0.007 to 2.8. 
 
 The comparison, however, assumes that our stellar mass statistics is reflected as a number statistic (resolved stars), a
 valid reasoning only if the stellar mass function does not depends strongly on metallicity.

 The detection of stellar haloes with low metallicity in resolved galaxies has been reported, for example,
 in NGC 3379 (\citealt{HAR07}) and in NGC 4486 itself. In this galaxy \citet{BIR10} find a low metallicity tail,
 using HST observations. However, the limiting magnitude in this last work prevents a reliable assessment 
 of the upper end of the metallicity distribution.  

 The presence of an extended stellar halo with low metallicity in NGC 4486 is also coherent with the detection of stars
 with ages older than $10$ Gyr and  metallicities below $[Fe/H]\approx -1.0$, at large distances from the dominant Virgo
 galaxies (\citealt{WILL07}).

\section{Large scale features: Up to 1000 arcsec (80.4 kpc) along the galaxy semi major axis.}
\label{Large}

 An extension of the previous discussion to a larger angular scale can be performed
 by combining the GC areal densities from \citet{HAR09} and the $(B-V)$ colour gradient of
 the galaxy presented by \citet{RUD10}.

 We start modelling  the spatial distribution of the GC sub-populations, and then look for the
 $\delta$ and  $\gamma$ that provide the best  fit to both the large scale colour gradient and
 of the galaxy profile brightness. Those parameters, in turn, will allow the determination of
 the chemical abundance distribution and (M/L) ratios for field stars, as well as of the total
 stellar mass.   

\subsection{Areal densities.}
\label{Areal}

 A usual approach in the literature is the adoption of $r^{1/4}$ or of power laws
 in order to fit the GC areal densities as a function, for example, of  galactocentric 
 radius. Both \citet{TAM06} and \citet{HAR09} present this kind of analysis. Alternatively, the
 combination of their data with those given by \citet{JOR09} and FFG07, and taking into
 account the slightly different magnitude limits in these works, shows that \citet{SER}
 profiles can provide adequate fits even including the central regions of the galaxy, where GCs
 display core-like density profiles.

 As a first step, we found the S\'ersic parameters (scale length $r_s$ and $n$
 index) that fit the density profiles along the galactocentric radius (i.e. within circular
 annuli) for both GC populations. However, as already noted by \citet{MCL94} the GC system rather
 follows the galaxy flattening.

 Taking this feature into account, we generated a model areal distribution which, in the
 case of the red GCs, follows the ellipticity of the galaxy halo ($\epsilon = 0.03$
 at $a = 5$ arcsec to $\epsilon = 0.35$ at $ a = 500$ arcsec). For the blue GCs we set a constant
 ellipticity ($\epsilon = 0.40$). This last procedure seems justified since, in the outer
 regions, both the galaxy and the blue GC population exhibit a similar ellipticity (see fig 4.
 in \citealt{TAM06}). Comparable values of the ellipticities of the red and blue GC distributions
 have been presented in the extensive work by \citet{STR11}.

\begin{figure}
\resizebox{1\hsize}{!}{\includegraphics{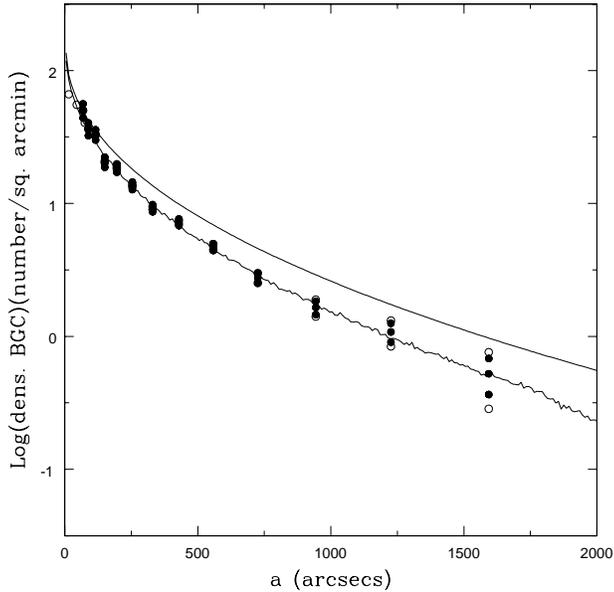}}
\caption{Blue GC areal density as a function of galactocentric radius (filled dots) from
 \citet{HAR09}, showing the statistical uncertainties of the GC counts. The effect of
 a variation of 50 per cent on the adopted background is represented by open dots.
 The upper solid line corresponds to a S\'ersic profile along the  semi major
 axis with constant ellipticity ($\epsilon = 0.40$) and characterized by $n=2.20$ and $r_s = 43.0 $
 arcsec. The lower wavy line, is the same model but sampled within circular annuli (i.e. within 
 constant galactocentric distances).
}
\label{figure9}
\end{figure}

\begin{figure}
\resizebox{1\hsize}{!}{\includegraphics{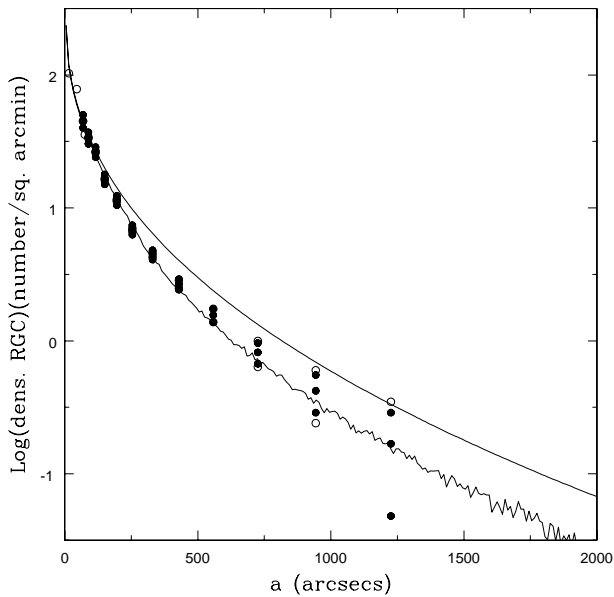}}
\caption{Red GC areal density as a function of galactocentric radius (filled dots) from
 \citet{HAR09}, showing the statistical uncertainties of the GC counts. The effect of
 a variation of 50 per cent on the adopted background is represented by open dots.
 The upper solid line corresponds to a S\'ersic profile along the galaxy semi major axis, following
 the galaxy ellipticity variation, and characterized by $n=2.35$ and  $r_s=13.5$ arcsec.
 The lower wavy line, is the same model but sampled within circular annuli (i.e. within 
 constant galactocentric distances).
}
\label{figure10}
\end{figure}

 The galactocentric (wavy lines) and semi major axis (continuous line) density runs of the models are shown
 in Fig. \ref{figure9} for the blue GCs ($r_{s} = 43$ arcsec, $n = 2.20$, central density= $170$ clusters per
 sq. arcmin) and  in Fig. \ref{figure10} for the red GCs ($r_{s}=13.5$ arcsec, $n=2.35$, central
 density = $350$ clusters per sq. arcmin). These diagrams also display the uncertainties of the profiles as
 a result of the adopted background and GC counting statistics.

\subsection{(B-R) Colours.}
\label{BMRcolours}

 In order to fit the integrated colours of the galaxy we followed the approach described in paragraph 3,
 adopting the GC areal density profile models to determine the number of blue and red GCs along the semi major
 axis (previous subsection).

 In turn, the variation of the GC chemical abundance scales within 300 arcsec were approximated  through linear
 relations: 

\begin{equation}
log(Z_ {SB}) = -0.25(\pm 0.02)~log(a) - 1.02(\pm 0.07)
\end{equation}
\noindent for the blue GCs, and
\begin{equation}
log(Z_{SR}) = -0.25(\pm 0.03)~log(a) + 0.55(\pm 0.07)
\end{equation}
\noindent for the red GCs, which are compatible with the abundance gradient derived from Table 1 (since
 $\Delta [Z/H]$/$\Delta$$log(Z_{s})= 0.69$ for both GC sub-populations). As a first approach these gradients
 were extrapolated outwards $a=$ 300 arcsec.

\begin{figure}
\resizebox{1\hsize}{!}{\includegraphics{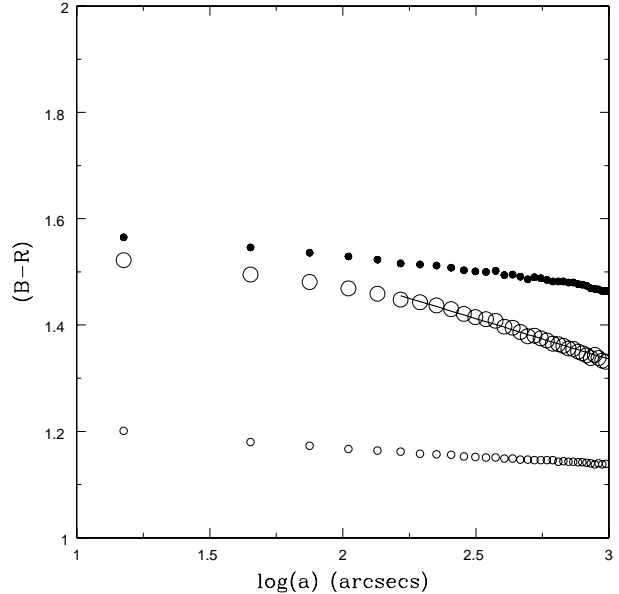}}
\caption{Logarithmic {\it (B-R)} colour vs. semi major axis $a$. Large open dots (0.01 mag in radius)
 are the model colours obtained for the galaxy and adopting $\delta=1.70$. The straight line is the colour
 gradient derived from Rudick et al. 2010 transformed to $(B-R)$. Filled dots represent the bulge colours
 while, open small circles, correspond to the halo colours, both inferred from the abundance gradients,
 discussed in subsection 5.2. 
}
\label{figure11}
\end{figure}

 The  $(B-R)$ colour gradients are displayed in Fig. \ref{figure11}, where large open dots represent the 
 best fit model corresponding to $\delta=1.70$, while the straight line is the (transformed) \citet{RUD10}
 colour gradient. 

 The $\delta$ parameter, in this case, is about 6 percent smaller than that derived in Section 4. The difference
 arises because the colour gradient from Rudick et al., $\Delta(B-R)$/$\Delta$$log(a)=-0.14 $, is slightly
 steeper than that derived from the \citet{MIC00} data for the innermost regions.

 Fig. \ref{figure11} also includes the colour gradients of the halo and bulge stars, both governed by the metallicity
 gradients discussed in Section 4. Although these gradients are rather similar, the shape of the colour-metallicity
 relation leads to a shallower colour slope for the halo, a behaviour shared by the blue GCs, and also found in other
 galaxies (e.g \citealt{FORB11} or \citealt{FAI11}).

 As an important result, Fig. \ref{figure11} indicates that most of the colour gradient of the galaxy arises from
 the luminosity weighted composition of the halo and the bulge, whose relative contributions change along the 
 semi major axis. The metallicity gradients of each sub-population seem to play a relatively minor role regarding
 the composite stellar colour gradient. For example, adopting null colour gradients outwards $a=300$ arcsec, yields
 $(B-R)$ colours that are only 0.015 mag redder at $a=1000$ arcsec.

 Finally, as seen in  Fig. \ref{figure11}, the model integrated colours keep within $\pm 0.01$ mag from the observed
 ones and do not seem to require a significant variation of the stellar age, as reported in the photometric study
 by \citet{LIU05}.

\subsection{The NGC 4486 Brightness profile.}
\label{HALO}
\begin{figure}
\resizebox{1\hsize}{!}{\includegraphics{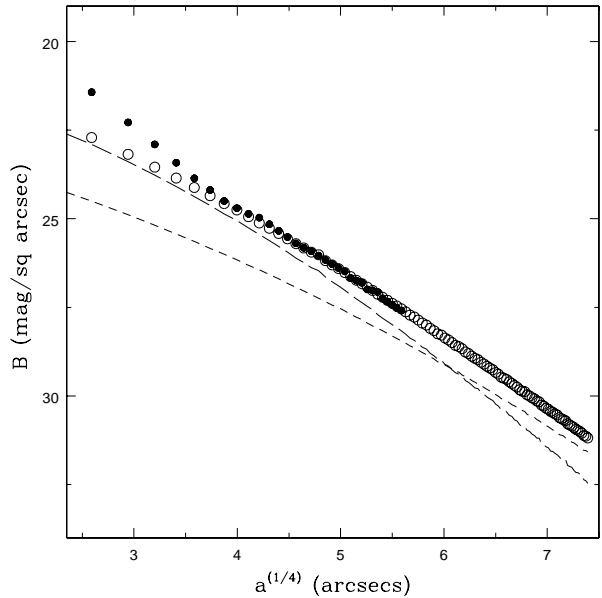}}
\caption{ NGC 4486 B surface brightness profile from  \citet{CAO90}, (dots) compared with
 the model profile derived using the GCs as stellar population tracers (open dots).
 The short and long dash lines correspond to the brightness contributions arising in the
 halo and bulge respectively, as traced by the blue and red GCs. The profiles are extrapolated
 between $a^{1/4}$=5.62, ($a=$1000 arcsec) and $a^{1/4}$=6.69, ($a=$2000 arcsec).
}
\label{figure12}
\end{figure}

  The \citet{CAO90} {\it B} profile adopted in this work, and shown in Fig. \ref{figure12}, is in   
  good agreement with previous ones (e.g. the E-W profile by \citealt{DEV78}, or the photographic
  work by \citealt{CAR78}), as can be seen in the comparison presented by \citet{LIU05}.

\begin{figure}
\resizebox{1\hsize}{!}{\includegraphics{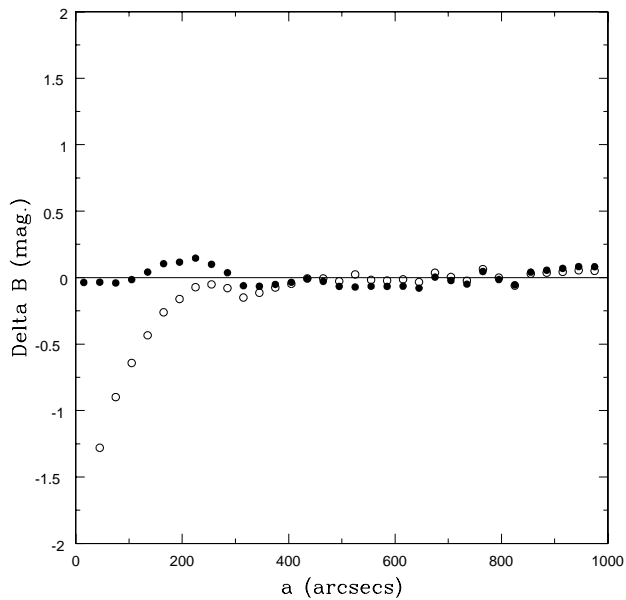}}
\caption{Brightness profile fit residuals using the GCs as luminosity tracers (open dots), compared
 with those arising from a single S\'ersic profile characterized by $n=10.0$. The
 strong deviation within $\approx 250$ arcsec is attributed to GC erosion by dynamical
 effects (see text).
}
\label{figure13}
\end{figure}

  A single S\'ersic profile fit  gives $n=10 \pm 2$, for a galactocentric range that covers from
  $a=20$ arcsec to $a=1000$ arcsec. The residuals, displayed in Fig. \ref{figure13},  correspond
  to  an overall $rms=\pm 0.07$ mag. The innermost region of the galaxy was not included to
  avoid the complex structure connected with the NGC 4486 jet as well as seeing effects. A similar fit,
  transforming  the \citet{CAO90} $B$ photometry to the {\it V} band, through the \citet{RUD10} $(B-V)$
  colours, leads to $n(V)=9.5 \pm 2$.

\begin{figure}
\resizebox{1\hsize}{!}{\includegraphics{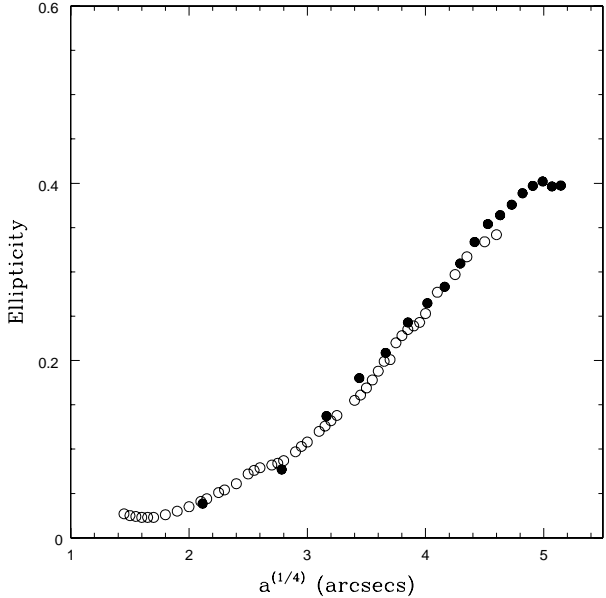}}
\caption{Ellipticity variation along the semi major axis of NGC 4486. Open dots are observed
values from Caon et al. (1990). Filled dots correspond to the bi-dimensional surface brightness
model described in the text.
}
\label{figure14}
\end{figure}

  Alternatively, we used  GCs as ``brightness tracers''. In this case, each ``diffuse'' stellar mass element
  (eq. 16) was spatially distributed following the description given in subsection 5.1 and, after adopting
  $\gamma=0.55\times10^{-8}$ and  $\delta=1.70$, transformed to $B$ luminosity to generate a bidimensional
  image of the galaxy. This image was analysed with the IRAF package to obtain the model surface brightness
  profile. The variation of the model  ellipticity along the semi major axis of the galaxy is compared
  with the observed one in Fig. \ref{figure14}. 

  The brightness residuals obtained by comparing the observed and model profiles, also depicted
  in Fig. \ref{figure13}, are very similar to those arising from the S\'ersic profile outwards
  $a=250$ arcsec.
  
  The systematic trend of these residuals towards the centre of the galaxy, is consistent with the behaviour
  of the $\gamma$ parameter shown in Fig. \ref{figure5}, and with the possible destruction of GCs by dynamical
  effects. Reconciling the observed brightness profile with that traced by the GCs, imply that we currently 
  observe 85 per cent and 60 per cent of the original blue and red GC populations, respectively.

  We stress that the completeness factors of the photometric works by  FFG07 and \citet{JOR09} are higher than
  90 percent for GC candidates brighter than the turn over of the GC luminosity function, then ruling out that
  this trend could be eventually caused by instrumental or galaxy background effects. 

  Our two components (halo and bulge) fit has some similarities with the approach presented by
  \citet{SEI07} who found that the brightness profiles of five cD galaxies are better
  represented by combining two S\'ersic profiles with different $n$ indices and scale lengths. 
  In particular, they noticed the common presence of outer exponential haloes ($n=1.0$). This result, in a
  first approximation, is not in conflict with ours since, forcing $n=1.0$ leaves brightness residuals
  smaller than $\pm 0.05$ mag for the halo component, within a range of $a=250$ arcsec to $a=1000$ arcsec.

  Fig. \ref{figure12} also suggests that the brightness of the extended halo should equal the bulge brightness at
  $a\approx 1300$ arcsec (104 kpc) and that, presumably, would become dominant outwards. This result is consistent,
  but better constrained than that in FFG07, mainly due to a more reliable definition of the colour gradient of
  the galaxy on an extended angular scale.


\subsection{A comparison between the chemical abundance of globular clusters and field stars.}
\label{Comparison}

\begin{figure}
\resizebox{1\hsize}{!}{\includegraphics{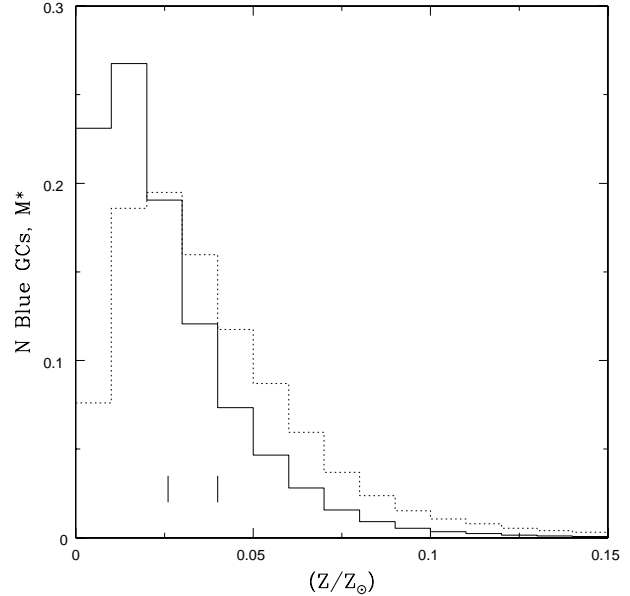}}
\caption{Chemical abundance distribution (in solar units) for the blue GCs within $a=1000$ arcsec (80.4 kpc)
 normalized by total number (solid histogram) and for the halo stars, normalized
 by  total mass (dotted histogram). Vertical lines at $(Z/Z_\odot)$=0.026 and $(Z/Z_\odot)$=0.040 indicate 
 the mean chemical abundance for clusters and stars, respectively.
}
\label{figure15}
\end{figure}

\begin{figure}
\resizebox{1\hsize}{!}{\includegraphics{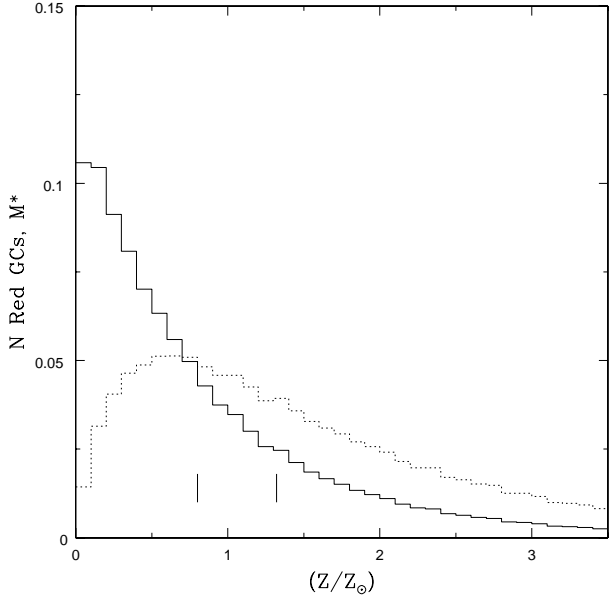}}
\caption{Chemical abundance distribution (in solar units) for red GCs within $a=1000$ arcsec (80.4 kpc),
 normalized by total number (solid histogram) and for the bulge stars, normalized by total
 mass (dotted histogram). Vertical lines at $(Z/Z_\odot)$=0.80 and $(Z/Z_\odot)$=1.32 indicate
 the mean  chemical abundance for clusters and stars, respectively.
}
\label{figure16}
\end{figure}

 The chemical abundance distribution function for the blue GCs and halo stars, as well as  
 for the red GCs and bulge stars, within $a=1000$ arcsec (80.4 kpc), are depicted in
 Fig. \ref{figure15} and  Fig. \ref{figure16}. These diagrams show  mean stellar abundances about 1.6  
 times (0.2 dex) larger than those of the GCs. 

 Preliminary interpretations of these mass-chemical abundance spectra can be made along the
 usual pathways, i.e., assuming that $Z$ is a ``clock'' as in simple models or, alternatively,
 that chemical abundance is rather a volume ``sampler'' as in the inhomogeneous enrichment
  models.

 A search of the literature shows that the inferred MDF of the bulge stars is
 rather similar to the ``push'' model put forward by \citet{SCH09}. These authors 
 artificially modify the output of an infall model, by reducing the low metallicity 
 end of the stellar population, in an attempt to asses the impact on the integrated
 model colours of elliptical galaxies  and  to avoid the {\it ``infamous G-dwarf problem'' }.

 They also reject the simplest chemical evolution scenarios and, at the same time, point
 out the similarity of their tentative approach with the results from inhomogeneous
 enrichment models  (\citealt{OEY00}) which also provide a better representation of the high 
 metallicity  drop-off of the MDFs. 

 The abundance distributions in the inhomogeneous enrichment models are the result of star
 formation in different events that eventually overlap, and are  are characterized  by the product $n.Q$,
 where $n$ is the number of stellar generations and Q is a filling factor that measures the relative
 volumes of enriched and primordial gas.

 In particular we note the similarity of the MDFs presented in this work and those 
 depicted in Oey's figures 2 and 3, corresponding to the halo and bulge of the MW, which require low 
 and high $n.Q$ values respectively. It is tempting to connect these results with the fact 
 that, the $\gamma$ and $\delta$ parameters derived in previous sections, imply that  $\approx$
 50 percent of the blue GCs form associated with diffuse stellar masses below $10^{7} M_\odot$,
 and with masses below $12\times10^{7} M_\odot$ in the case of the red GCs. 

 Another intriguing result is suggested by Oey's fig 1 (panel c).That diagram shows the decoupling 
 of the MDFs characterizing the very early stellar generations from the composite one at the
 end of the whole star forming process. In this scenario, the first stellar generations still reflect
 the so called ``metal production function'' $ f(Z), $ that is assumed to be a power law (decreasing with Z) 
 governed by SNe enrichment. In our analysis, GCs follow a comparable trend (although with an 
 exponential dependence) and could be in fact mapping such $f(Z)$  if their formation  is restricted to 
 the initial stellar generations.  

 In either case, the landscape seems coherent with \citet{SPIT10}, who argues in favour of an intense 
 formation of bulge-GCs followed by an era of stellar bulge growth and little cluster production.   


\subsection{Stellar mass to luminosity  ratio.}
\label{RML}

 The dependence of the mass to {\it B} luminosity ratio with chemical abundance
 adopted in this work (eq. 17) was used to determine integrated ratios
 for galaxies in the Virgo cluster (FFV09). The ratios inferred in that work are
 in perfect agreement with those reported by \citet{NAP10} through a completely
 different approach.

\begin{figure}
\resizebox{1\hsize}{!}{\includegraphics{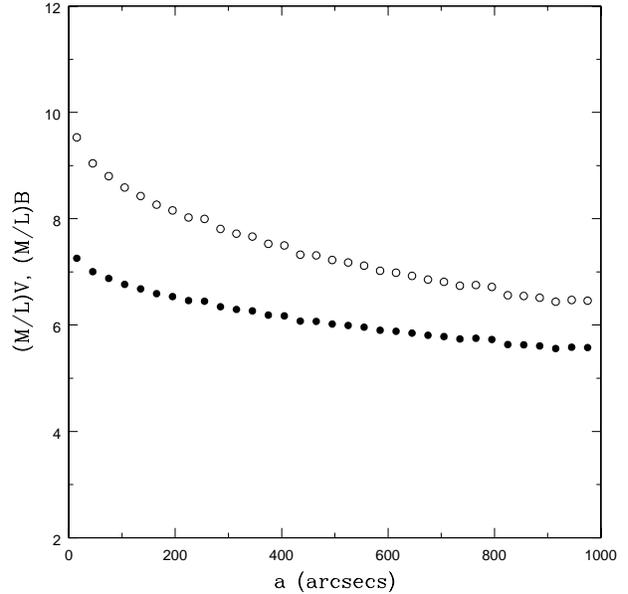}}
\caption{Inferred Blue (open dots) and Visual (filled dots) mass to luminosity ratios
 within 1000 arcsec (80.4 kpc) along the NGC 4486 semi major axis and within $30$ arcsec annuli
 following the variation of the ellipticity of the galaxy.
}
\label{figure17}
\end{figure}

 Using the same dependence, the chemical gradients described before, and adopting
 $\delta=1.70$ (subsection 5.2), we derived the mass to luminosity ratios for
 the {\it B} and {\it V} bands (within elliptical annuli $30$ arcsec wide in $a$) for the
 composite stellar population, i.e., including both halo and bulge stars, that are 
 plotted in Fig. \ref{figure17} and correspond to an overall
 integrated ratio  $(M/L)_{B}=8.3$.

 Most of the variation of the $(M/L)$ ratios in this last figure, as in the case of the colour
 gradient of the galaxy, is governed by the change of the mass ratio of halo to bulge stars 
 with galactocentric radius. Halo stars alone, in fact, show a rather limited range of the
 $(M/L)_{B}$ ratio, which falls from 4.4 to 4.2 within the inner $120$ arcsec and levels off at
 $ \approx$ 4.0, outwards.

 Recently, \citet{GEB09} and \citet{MUR11} made an extensive analysis of the stellar
 and GC kinematics in NGC 4486. Their dynamical model includes  a super massive
 black hole, stars and dark matter. The $(M/L_{V})$ ratio shown in Fig. \ref{figure17}  
 is comparable to $(M/L)_{V}=6.3 \pm 0.8$, given in the first paper, but significantly 
 lower than the range $(M/L)_{V}= 8.2$ to $9.1$ reported in the second work.
 
\section{Stellar and dark matter masses.}
\label{stellardark}

\begin{figure}
\resizebox{1\hsize}{!}{\includegraphics{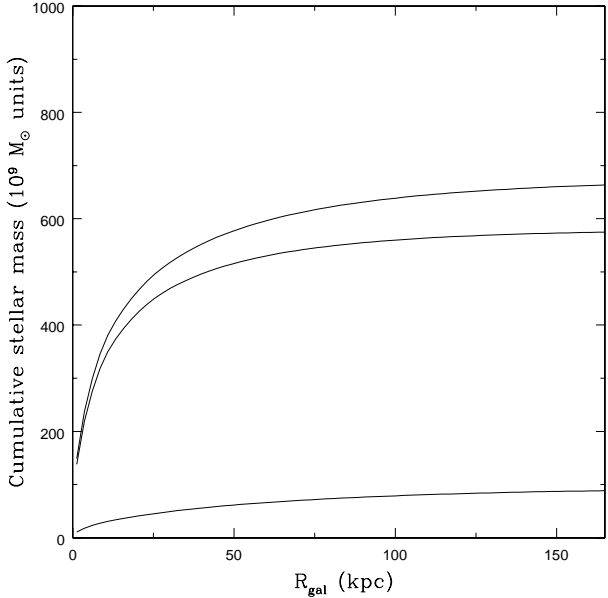}}
\caption{Cumulative stellar mass as a function of galactocentric radius for the
 total (upper curve), bulge (middle curve) and halo (lower curve) stellar populations.
 The half (projected) mass radii for the bulge and the halo are $ 8.4$ Kpc ($105$ arcsec) and
 $28$ kpc ($350$ arcsec) respectively.
}
\label{figure18}
\end{figure}

  The halo, bulge and total (projected) cumulative masses are shown in Fig. \ref{figure18}.
  These curves were obtained by combining the Caon et al. photometry and the $(M/L)_{B}$
  ratios discussed in the previous section within $a=1000$ arcsec. Outwards, where the 
  stellar mass contributes with only  $\approx 10$ percent of the total, we assumed constant ratios
  set by the outermost $(M/L)_{B}$ value. 

  Using kinematic data of planetary nebula, \citet{DOH09}, determine the ``edge'' of the NGC 4486 halo
  to be at $\approx$$2000$ arcsec (164.8 kpc), which is the adopted upper galactocentric limit in Fig. \ref{figure18}. 
  Within this galactocentric radius we obtain a total stellar mass of  $6.8(\pm 1.1)\times10^{11} M_\odot$.
  This mass is consistent the S\'ersic index $n_{V}$ vs-stellar mass relation derived by \citet{DON11}
  (see their fig. 1, upper left panel) in an analysis  of the connection between the shape of early type 
  galaxies and their stellar populations. 

  Our estimate of the mass uncertainty, about $\pm$ 15 percent, comes from errors of $\pm$ 0.05,
  and $\pm$ 0.10, for the $\gamma$ and $\delta$ parameters, respectively, and of $\pm$ 0.25 mag for the
  integrated blue magnitude of the galaxy ($B=$9.50). 

  Fig. \ref{figure18} indicates that the halo and the bulge make different contributions to the total stellar mass
  (14 per cent and 86 per cent, respectively) and also that these subsystems have very distinct spatial distributions
  characterized by half mass projected radii of $110$ and $350$ arcsec (8 and 28 kpc, respectively).

   A similitude between the projected blue GCs areal densit given by FFG07 and that of dark matter in NGC 1399 (from
  \citealt{RICH04}), was  pointed out in \citet{FFG05}, and still holds after the  recent and improved analysis by
  \citet{SCHU10} (see their models a10 and b10). The same comparison can be performed for NGC 4486 using
   the results presented by \citet{GEB09} and \citet{MUR11}. 

   In this case, and even though the total enclosed mass derived by those authors are in good agreement
  ($2$ to $3\times10^{13} M_\odot$ within $160.8$ kpc), their inferred spatial distributions for 
  dark matter are significantly different. Under the assumption of a logarithmic potential to
  represent the dark halo, those works obtain core radii in the range from $164$ arcsec to $414$ arcsec.  

\begin{figure}
\resizebox{1\hsize}{!}{\includegraphics{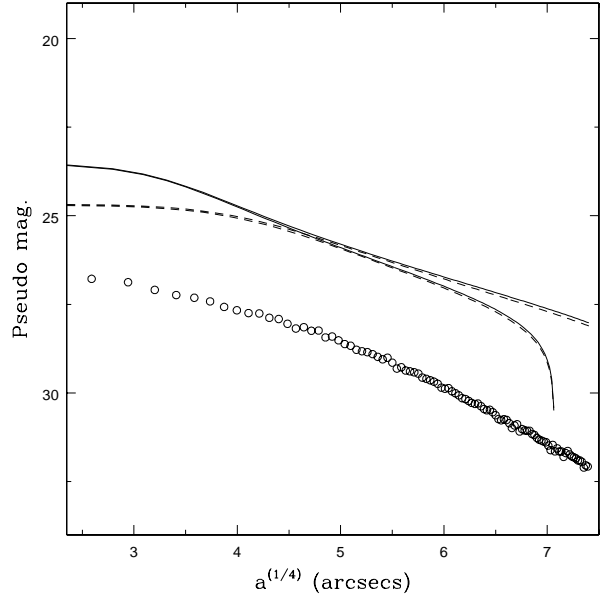}}
\caption{A comparison between the projected dark matter density profiles (on a pseudo
 magnitude scale) from Gebhardt and Thomas (2009) (continuous line) or Murphy, Gebhardt and Adams (2011)
 (dashed line),  with the inferred surface brightness of stars with $(Z/Z_\odot)$ less than 0.025
 (open dots), along the NGC 4486 semi major axis. The projected dark matter profiles correspond to cut off
  radii of 2500 arcsec (lower curves) and 5000 arcsec (upper curves) respectively.
}
\label{figure19}
\end{figure}

  The corresponding dark matter volumetric density profiles (eq. 3 in \citealt{GEB09}), once projected
  on the sky and given on a pseudo magnitude scale  arbitrarily shifted in ordinates, are displayed in
  Fig. \ref{figure19}. As the outer spatial limit of the dark 
  matter halo is not well constrained we adopted two tentative values in order to asses the impact on 
  the projected profile. The lower curves correspond to a cut off radius of $2500$ arcsec, i.e., 25 per cent 
  larger than that given by \citet{DOH09}, while the upper curve corresponds to twice that radius.

  This figure also shows the inferred surface brightness profile  of the stellar population with a chemical
  abundance lower than $(Z/Z_\odot)=0.025$. These astars are connected with GCs {\bf{bluer}} than the ``blue peak'' (at
  $\approx (C-T_1)=1.25$) of the cluster colour distributions.

  A meaningful comparison closer to the centre of the galaxy (i.e. within $250$ arcsec) becomes blurred  by
  dynamical erosion on the blue GCs, used as tracers of the stellar component, and would require a rigurous
  assessment of these effects.
  
  Even though  dark matter and  very low metallicity stars display rather similar projected distributions, 
  a comparison between their spatial distributions still remains somewhat ambiguous due to the uncertain cut off 
  radius of the halo. On one side, adopting the \citet{DOH09} value, would suggest that these stars in fact ``map''
  the dark halo. Alternatively,  doubling that cut off radius, leads to a somewhat shallower distribution for
  dark matter. In fact, \citet{STR11} do not find evidence for a transition in the inner halo to a potential
  dominated by the Virgo cluster. 

  A shallower distribution of the dark matter is predicted by the \citet{ABA06} models for isolated galaxies,
  where an extended  halo develops by accretion of stars shed from early merging sub-units, in an hierarchical
  galaxy formation scenario. A further and rigorous comparison, however, should take into account the distinct
  environmental situation of NGC 4486 as a central galaxy in the Virgo cluster.
 
  The results discussed in this subsection also allow an estimate of the baryon fraction ($f_b$) in
  NGC 4486, which we restrict to a galactocentric radius of $80.4$ kpc aiming at avoiding the inherent
  uncertainties that become more important at larger radii. The total baryonic mass, within that
  boundary, and in the form of stars, is $6.8\times10^{11} M_\odot$, while the hot gas amounts to
  $2.0\times10^{11} M_\odot$, from \citet{FAB83} (and after correcting the distance adopted by
  these authors, $d=15$ Mpc, to ours). On the other side, the total enclosed mass in both \citet{GEB09} and 
  \citet{MUR11} is close to $10^{13} M_\odot$, then yielding a fraction $f_b=0.08$. This value is
  well below the cosmic baryon fraction (0.17). However, combined with the circular velocities
  (715 to 800 km/s) given in these last two works, it falls close, and within the dispersion, of 
  the baryonic fraction-vs-circular relation presented by \citet{DAI10} (see their figure 4).   


\section{ Globular cluster formation efficiency.}
\label{EFFIC}

 The estimated number of blue and red globulars within $a = 2000$ arcsec, assuming a fully Gaussian
 GC integrated luminosity function, (12200$\pm$ 1500 and 4600$\pm$ 1000) corresponds to an specific $S_n$
 frequency close to $14 \pm 1.5$ and is consistent with the already known nature of NGC 4486 as an
 archetype ``high  $S_n$ system".

\begin{figure}
\resizebox{1\hsize}{!}{\includegraphics{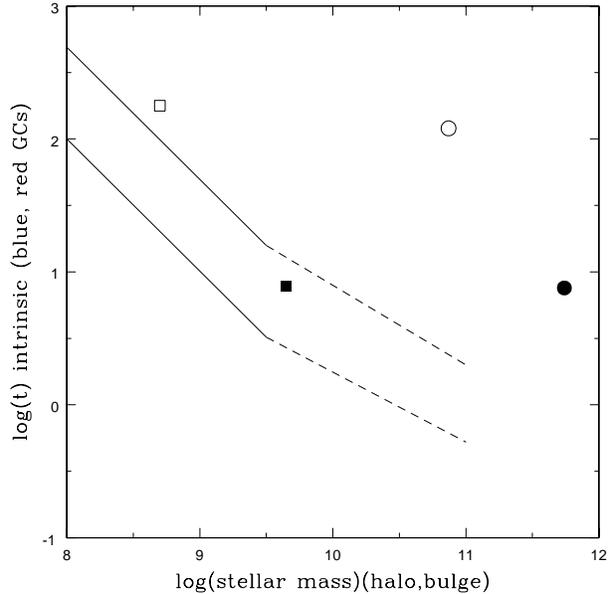}}
\caption{Globular cluster intrinsic formation efficiency locus for the blue (solid
 lines) and red (dash lines) sub-populations associated with 63 galaxies in the Virgo cluster. 
 The open and solid dots represent the integrated values of the blue and red GCs associated
 with NGC 4486, respectively. Squares correspond to the MW GCS. 
}
\label{figure20}
\end{figure}

 The number of GCs, and the halo and bulge masses discussed in Section 6, yield the {\bf{intrinsic}}
 GCs formation efficiency for each subsystem in terms of the associated halo mass $log(t_{B})= 2.15$, 
 or bulge mass, $log(t_{R})=0.90$ (both given in $10^{-9}$ units).
 These values are  shown in Fig. \ref{figure20} and compared with the locus corresponding
 to GCS in Virgo galaxies, derived from a re-discussion of the FVF09 data. In this last work,
 we note, the definition of GC formation efficiency is  that given by \citet{ZEP93}, i.e.,
 in terms of the total stellar mass of the galaxy (see also \citealt{RHO05}). 

 Fig. \ref{figure20} also shows the MW GC formation efficiencies, after adopting the halo and bulge masses
 from the discussion presented by \citet{BOL09}, and a number of 110 halo and 35  bulge GCs (\citealt{MAC05};
 \citealt{BIC06}). The MW clusters also show a much higher formation efficiency for the
 halo GCs, and compare well with the Virgo GCS locus which, in principle, seems dominated by galaxies that
 have suffered minor merging events. Even though the GC formation efficiencies presented in this work depend
 on the assumed connection between the number of GCs and the diffuse stellar mass (Section 3), those of the
 MW GCs  come from a completely independent approach.
 
 The reason behind the strong difference in the formation efficiency of halo and bulge 
 globulars is not clear. However, a distinctive feature between both systems,
 besides metallicity, seems to be the mean density of the associated stellar populations.
 A first estimate of these densities, using the half mass radii and stellar masses given in the
 previous subsection, indicates a mean bulge stellar density  an order of magnitude higher than
 in the halo.

 In an speculative way, this suggests that the erosion of potential GC formation sites by field stars may
 be more important in the bulge environment then leading to a lower cluster formation efficiency.
 
 In turn, Fig. \ref{figure20} indicates that dry merging might have provided both the halo stellar
 mass and the high number of blue GCs in NGC 4486 through some 300 galaxies with indicative masses
 of $\approx 30\times 10^{7} M_\odot$.  

 Within this frame, and as a tentative approach, we created a ``composite'' galaxy using
 the \citet{JOR09} data corresponding to ten Virgo galaxies with more than fifty blue GCs
 each.

 As some works indicate that galaxies can reach  masses $\approx  10^{10.8} M_\odot$
 before the star forming process is self interrupted (see \citealt{KHO09}), this limit was adopted
 to define the upper mass of the galaxies in the sample: VCC 685, 698, 759, 1062, 1231, 1279, 
 1297, 1327, 1692, and 2000. That value is close to $10^{11}  M_\odot$, upper mass
 of the galaxies that define the 3-D plane discussed in FVF09.

 The total population of the composite galaxy includes $916$ GC candidates brighter than the
 NGC 4486 GCs turn-over magnitude ($g=23.87$ mag; \citealt{VIL10}). 
 
 From the analysis of this data set we emphasize:

\noindent a) The composite GCs colour-magnitude diagram shows a blue tilt similar to the
 tilt observed for the blue GCs in NGC 4486.

\noindent b) A fit similar to those given in Table 1, leads to GC chemical
 scale lengths of $Z_{SB} = 0.015$ and $Z_{SR} = 0.55$.
 
\noindent c) The blue GCs chemical scale parameter, $Z_{SB}$, is comparable to that expected
 for the blue GCs in the outer regions of NGC 4486.

\noindent d) The red GC colours can be adequately fit by a single $Z_{SR}$ parameter even though they  
 are the result of composing galaxies that exhibit a large range in their individual parameters
 (0.12 to 0.90).

 These results give support to the idea that the blue tilt is a generalized feature in most
 Virgo galaxies and could  be detected even after eventual merging processes. 

 Although the accretion of low mass galaxies might have provided an important fraction of the NGC 4486 halo
 mass, as suggested in the {\it naive} approach by \citet{FOR82}, the presence of a metallicity 
 gradient for the blue GCs rather indicates that dissipative collapse has also played a role.
 
 On the other side, dry merging alone cannot explain the characteristics of the red GCs-bulge in 
 NGC 4486 since the inferred $Z_{SR}$ parameters in this galaxy reach values that are as twice 
 as those of the individual galaxies with comparable $t_{R}$ in Fig. \ref{figure20}. The 
 result commented in item $d$ is also interesting in the sense that eventual multiple mergers
 will not yield detectable effects on the (almost always) bimodal colour distributions.
 
\section{Conclusions}
\label{CONCLU}

 The distribution of the integrated GC colours in NGC 4486 is compatible with the
 presence of two different (blue and red) populations both following an exponential
 decay (in number) with increasing chemical abundance Z. The chemical scale
 lengths, however, are very different, being much smaller for the blue GC
 family. Both GC chemical scales have negative gradients with increasing
 galactocentric radius. In turn, the composite mass weighted chemical gradient obtained
 for the composite stellar population is comparable to that derived through X ray observations
 of hot gas, indicating a strong connection between stars and the interstellar medium.

 The main idea behind the GCs-field stars connection adopted in this
 work is that the metallicity of a given cluster is a proxy to the mass of
 the associated diffuse stellar population where it is born and which, through
 pre-enrichment (see \citealt{BAI09}), determines the cluster metallicity 
 itself. Furthermore, metallicity may be indicative of the mean stellar
 density of the environment, a factor possibly governing the GCs formation
 efficiency.

 The approach allows, by combining both the halo and bulge components, the reconstruction
 of the galaxy profile brightness and of the negative integrated colour gradient over
 an angular scale of, at least, $1000$ arcsec ($\approx$ 80 kpc). Given the multi-metallicity
 nature of the stellar population, SSP models are not adequate for a proper understanding
 of integrated colours in terms of chemical abundance (for a detailed discussion on this
 subject see \citealt{SCH09}).

 The chemical abundance distribution inferred in this work for the stellar population is similar to 
 those observed in resolved galaxies and exhibits similarities with the predictions given by heterogeneous 
 enrichment models as discussed by \citet{OEY00}. 

 In particular, the halo/bulge systems can be identified as formed in low/high density environments 
 typified by small/large number of stellar generations and different filling factors.
 A comparison of our Fig. 15 and 16 with figs. 2 and 3 in Oey's work, appeals  for a further, more
 quantitative exploration.

 If GCs appear at the very beginning of the star forming process their own chemical distribution
 may be reflecting the original ``metal production function'' and this would also imply larger mean
 ages than that of the field stars. In agreement with this, \citet{SPIT10} suggests that metal
 rich GCs should be older by about $1$ Gyr compared with the average age of the bulge.

 The GC formation efficiency for each (halo, bulge) system shows that the NGC 4486 system clearly
 deviates from those corresponding to other Virgo galaxies with masses smaller that $10^{11}  M_\odot$.

 Although both the NGC 4486 stellar halo mass, and the large number of blue globulars, could be reached
 by accreting low mass galaxies, the presence of a chemical abundance gradient rather suggests that
 accretion might have coexisted with the in-situ formation of indigenous blue GCs (see \citealt{HAR09} or
 \citealt{FORB11} for a similar conclusion). A description of the  mechanisms that may play a simultaneous
 role in the formation of galaxy haloes is given, for example, in the model analysis by \citet{MUR07}.

 We also stress that the eventual accretion of an external population of blue clusters, through numerous small
 mergers, would not affect the detectability of the chemical or colour gradient of the indigenous blue GCs  or of
 the blue tilt (as shown in Section 7). Alternatively, but not in necessary disagreement, \citet{KISS98} argue that,
 their detection of rotation for the blue GCs, is possibly the fingerprint of an important single past merger.  
 
 On the other side, pure dry merging cannot explain the red GC-bulge population since
 the chemical scale parameter of the plausible progenitor galaxies is much smaller
 than that in the central region of NGC 4486. Both dissipative collapse and/or wet
 merging leading to high star forming efficiencies and chemical enrichment seem required.
 The low central surface brightness, as well as the shallow chemical gradient in NGC 4486
 are compatible with merging processes (e.g. \citealt{KIM10}). This is also suggested by the
 position of the galaxy in the S\'ersic $n$ index-vs-central brightness relation ``normal'' ellipticals 
 (see \citealt{GRA03} for a comprehensive discussion). A scenario were GC formation is in fact entirely
 merger-driven is given in \citet{MURG10}. 

 The spatial distribution of the halo stars with the lowest chemical abundance is remarkably similar to that
 inferred for dark matter. This coincidence was already noticed in the case of NGC 1399 (\citealt{FFG05}).
 This last galaxy and NGC 4486 seem to have a very similar baryonic mass (see \citealt{JON97}) and also a
 comparable number of red GCs. However, they differ both in the number of blue GCs and dark mass content
 by the same factor, which is $\approx 3$ larger in NGC 4486 . Besides, a re discussion of the photometric data
 presented in FFG07 and of the dark matter content presented by Schuberth et al. (2010) for NGC 1399, also indicates 
 a remarkably similar number of blue GCs per dark mass unit, $\approx$ $1.0(\pm 0.3)\times10^{-9}$ in both
 galaxies.

 All these features add to the discussion of the possible origins of GCs, and in particular
 of the blue clusters, in connection with primordial dark matter sub-haloes (e.g. \citealt{SAN05};
 \citealt{MAS05}; \citealt{BEK08}; \citealt{BOL09}). 

 Further arguments about the connection between GCs and dark matter come from the
 very large scale study by \citet{LEE10}, who report the detection of ``intracluster''
 GCs whose projected spatial distribution resembles those of dwarf galaxies and dark matter
 on very large angular scales. The change in the slope of the GC areal density profile in their
 work, at $\approx$ $40$ arcmin (193 kpc) from the centre of NGC 4486, is close to the ``edge'' of 
 the galaxy halo as  determined by \citet{DOH09} (but see \citealt{STR11} for a different conclusion).

 The analysis presented in this work strongly support the idea that GCs appear along the main star
 forming events in early type galaxies and can be used as tracers of the dominant stellar populations.
 The ``two phases'' process, as suggested by \citet{FORB97}, seems an adequate scenario to explain the
 overall features of GCs and associated stars. The meaning of ``phase'', however, may require further
 clarification in terms of the existence of a temporal sequence (i.e. halo globulars appear
 prior to to bulge globulars as suggested by the very different chemical abundances ?), environmental 
 differences (underlying in the inhomogeneous enrichment models) and, eventually, about how both factors
 combine in the overall landscape of galaxy formation. 

\section*{Acknowledgments}
     
     JCF acknowledges Dr. Stephen Strom for introducing him, thirty years ago, to the fascinating NGC 4486 GC system,
     and also the Osservatorio Astronomico di Bologna and  Dr. A. Buzzoni  for their hospitality during 
     the last stages of this paper. This work was supported by grants from La Plata National University, and CONICET 
     (PIP 712), Argentina. FF acknowledges partial financial support from the Agencia Nacional de
     Promoci\'on Cien\'ifica y Tecnol\'ogica (BID AR PICT 885).

\appendix
\section{}

\begin{figure*}
\resizebox{1.0\hsize}{!}{\includegraphics{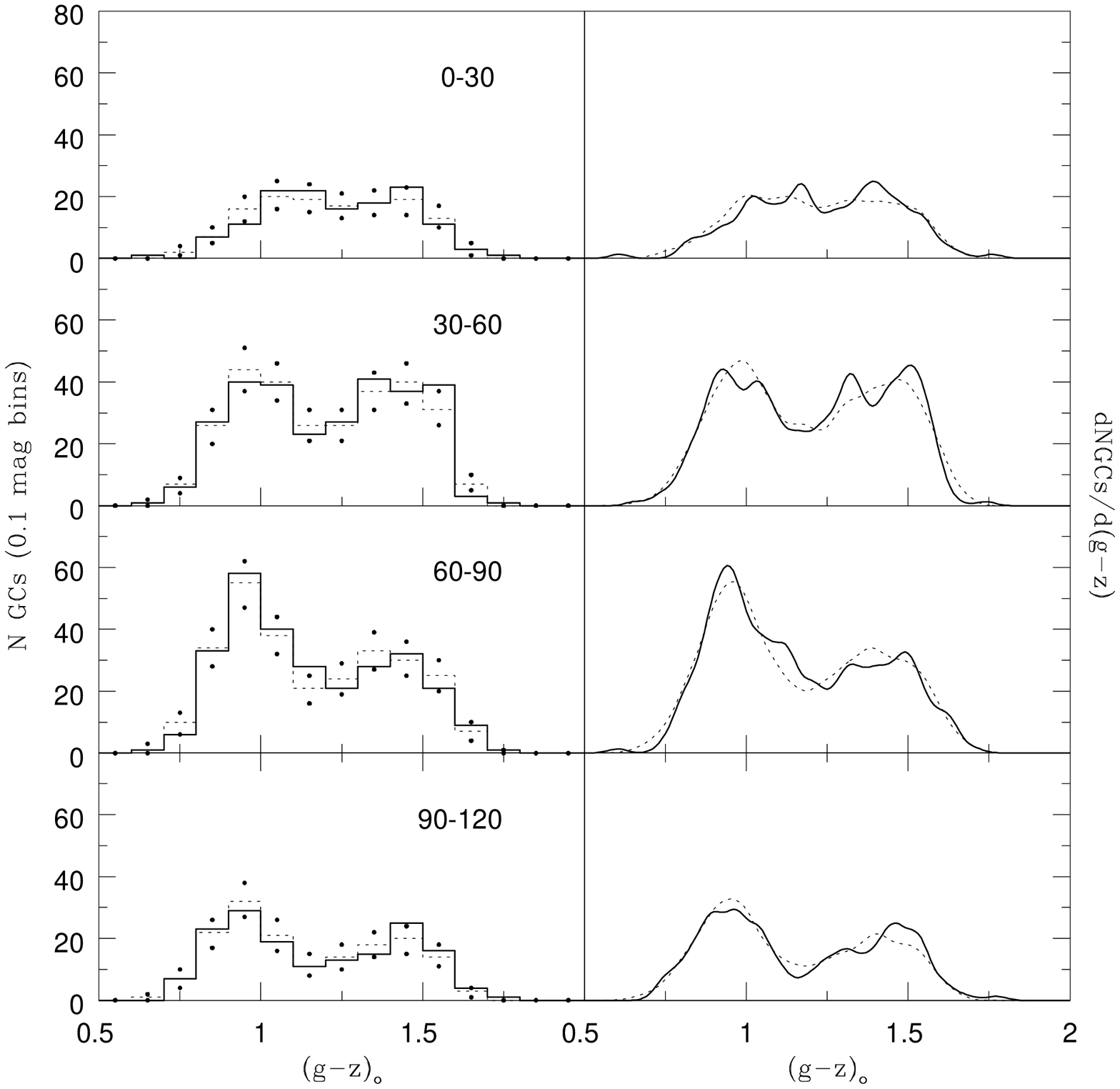}}
\caption{Left panel: Observed (continuous line) and model GC (g-z)o colour histograms adopting
 0.1 mag bins. Dots indicate the statistical uncertainty of the model within each bin. Right panel: Smoothed histograms
 adopting a Gaussian 0.03 mag kernel. The mean semi major axis of the annuli are given in arcsec. 
}
\label{figure20b}
\end{figure*}

\begin{figure*}
\resizebox{1.0\hsize}{!}{\includegraphics{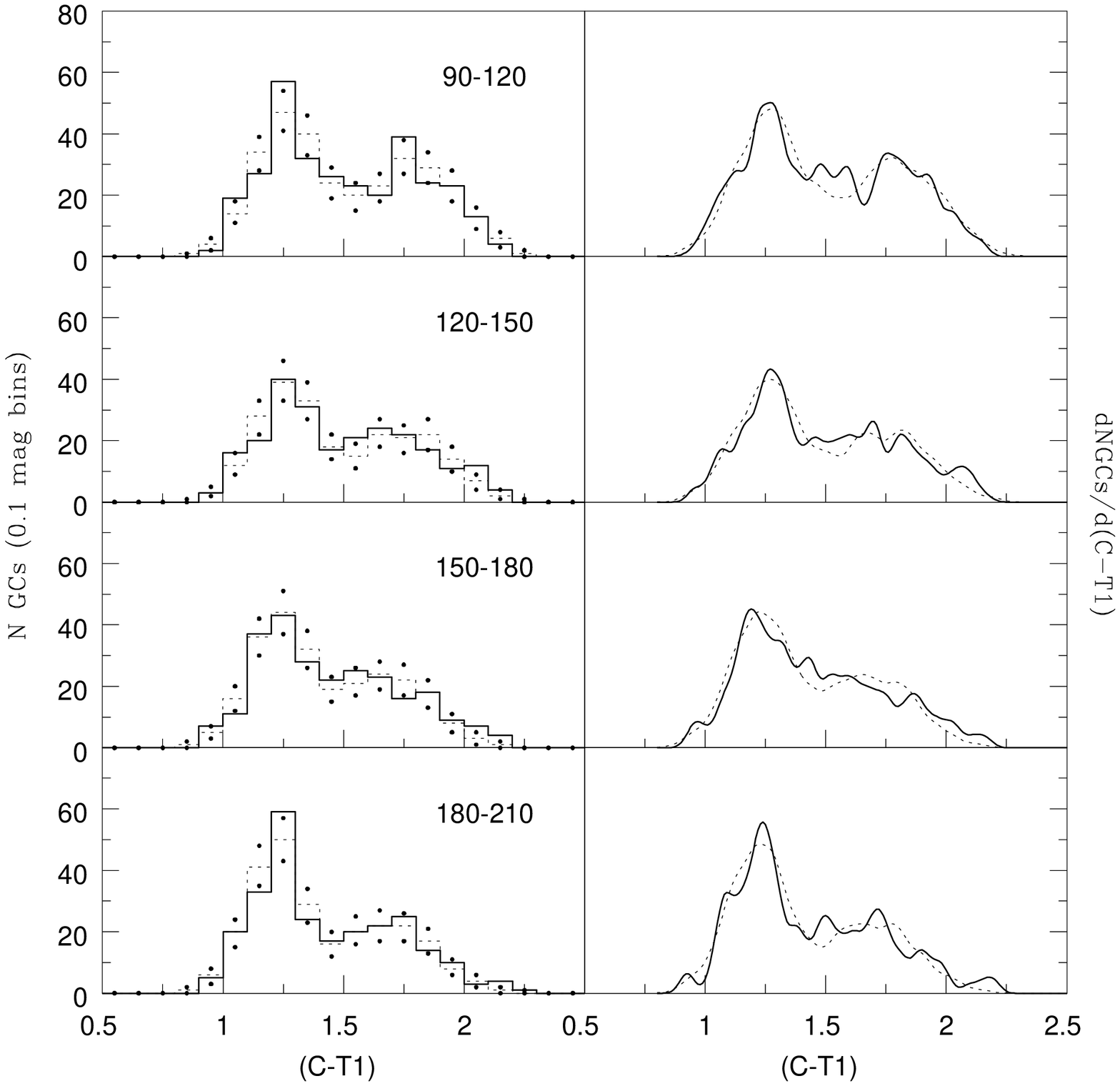}}
\caption{Left panel: Observed (continuous line) and model GC $(C-T_1)$ colour histograms adopting
 0.1 mag bins. Dots indicate the statistical uncertainty of the model within each bin. Right panel: Smoothed histograms
 adopting a Gaussian 0.03 mag kernel. The mean semi major axis of the annuli are given in arcsec. 
}
\label{figure21}
\end{figure*}

\begin{figure*}
\resizebox{1.0\hsize}{!}{\includegraphics{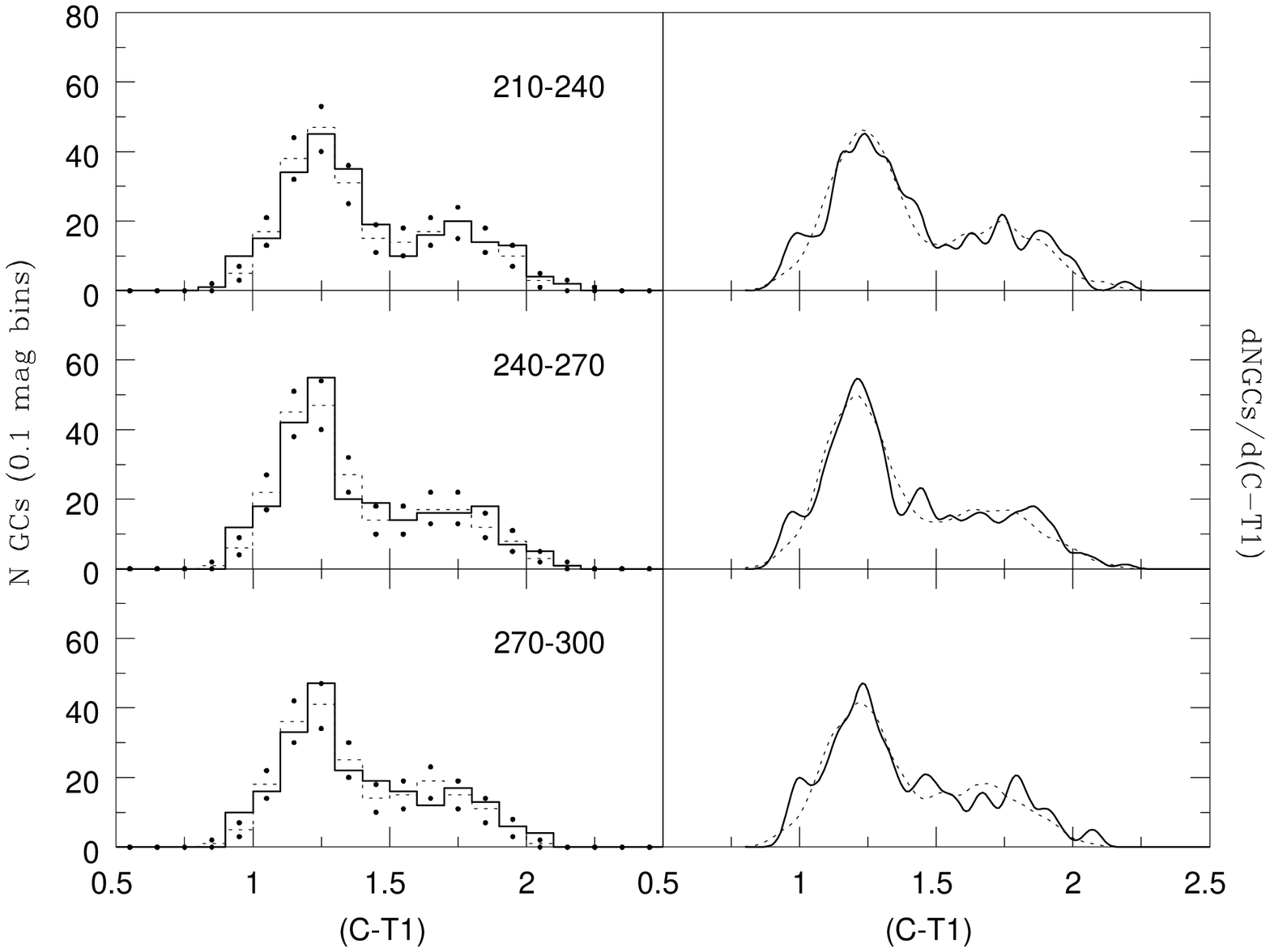}}
\caption{Left panel: Observed (continuous line) and model GC $(C-T_1)$ colour histograms adopting
 0.1 mag bins. Dots indicate the statistical uncertainty of the model within each bin. Right panel: Smoothed histograms
 adopting a Gaussian 0.03 mag kernel. The mean semi major axis of the annuli are given in arcsec.
}
\label{figure22}
\end{figure*}

\label{lastpage}
\end{document}